\documentclass[%11
superscriptaddress,
 amsmath,amssymb,
prl,
floatfix,
twocolumn
]{revtex4-2}

\usepackage{graphicx}% Include figure files
\usepackage{dcolumn}% Align table columns on decimal point
\usepackage{bm}% bold math
\usepackage{booktabs}
\usepackage{multirow}
\usepackage{braket}
\usepackage{xcolor}
\usepackage{makecell}
\usepackage[colorlinks]{hyperref}
\usepackage{pifont}
\hypersetup{
citecolor={blue},
urlcolor={blue}
}

\begin{document}

\preprint{APS/123-QED}

\title{
Macroscopic Polarization and Magnetization from Cavity Vacuum Fluctuations
}% Force line breaks with \\

\author{Jingkai Quan}
\email{jingkai.quan@mpsd.mpg.de}
\affiliation{%
Max Planck Institute for the Structure and Dynamics of Matter,
Luruper Chaussee 149, 22761 Hamburg, Germany
}%
\author{Chongxiao Fan}%
\email{chongxiao.fan@mpsd.mpg.de}
\affiliation{%
Max Planck Institute for the Structure and Dynamics of Matter,
Luruper Chaussee 149, 22761 Hamburg, Germany
}
\affiliation{Institute for Theory of Statistical Physics, RWTH Aachen University, and JARA Fundamentals of Future Information Technology, 52062 Aachen, Germany}
\author{Benshu Fan}
\email{benshu.fan@mpsd.mpg.de}
\affiliation{%
Max Planck Institute for the Structure and Dynamics of Matter,
Luruper Chaussee 149, 22761 Hamburg, Germany
}%
\author{I-Te Lu}
\affiliation{%
Max Planck Institute for the Structure and Dynamics of Matter,
Luruper Chaussee 149, 22761 Hamburg, Germany
}%
\author{Dante M. Kennes}
\affiliation{Institute for Theory of Statistical Physics, RWTH Aachen University, and JARA Fundamentals of Future Information Technology, 52062 Aachen, Germany}
\affiliation{%
Max Planck Institute for the Structure and Dynamics of Matter,
Luruper Chaussee 149, 22761 Hamburg, Germany
}%
\author{Angel Rubio}
 \email{angel.rubio@mpsd.mpg.de}
\affiliation{%
Max Planck Institute for the Structure and Dynamics of Matter,
Luruper Chaussee 149, 22761 Hamburg, Germany
}%

\affiliation{Initiative for Computational Catalysis, The Flatiron Institute, Simons Foundation, New York City, NY 10010, United States of America}

\date{\today}% It is always \today, today,

\begin{abstract}

Cavity light-matter interaction has recently emerged as a new avenue for manipulating material properties without driving fields. Here, we demonstrate that cavity vacuum fluctuations can induce macroscopic polarization (magnetization), even in materials that lack spontaneous polarization (net magnetization) in free space. Starting from the effective photon-free quantum-electrodynamics Hamiltonian, we identify all crystallographic (magnetic) point groups that allow such cavity-induced responses. We derive the form of the corresponding response tensors based on symmetry analysis, whose elements can be obtained by quantum electrodynamical density functional theory (QEDFT) calculations. As representative examples, we show that the cavity-induced polarization in $\alpha$-quartz can be continuously controlled by rotating the cavity. For antiferromagnetic Mn$_3$Sn, we demonstrate that cavity-induced symmetry breaking generates an out-of-plane magnetization, accompanied by an anomalous Hall conductivity component that is forbidden outside the cavity.
Our work establishes symmetry as a guiding principle for cavity materials engineering and provides a route for controlling polarization and magnetization through quantum vacuum fluctuations, i.e., cavity materials engineering.
\end{abstract}

\maketitle

%\tableofcontents

% \section{Introduction}
{\it Introduction---}Manipulating the electric and magnetic properties of materials is of fundamental interest in condensed matter physics and materials science~\cite{ferroelectric_rmp, spintronic_Nobel_rmp, Tokura_multiferro_review, Keimer_Moore_qm_review, quantum_materials_roadmap}. 
Beyond conventional approaches based on electric and magnetic fields~\cite{mag_control_pol_nature, elec_control_mag_rmp}, strain~\cite{piezo_em_review_AEM,piezoelectric_review, strain_engineering_review}, or chemical modification~\cite{doping_em_am},
light-matter coupling offers a new route to tailor these properties~\cite{Oka_floquet_review, ultrafast_rmp_2021,Disa_light_engineer_review,Cavalleri_nonlinear_phonon_review, ultrafast_NRM_2023,peizhe_ultrafast_NRP,angel_nature_perspective, schlawin2022cavity,cavity_review_2026}.
In particular, 
cavity materials engineering has recently emerged as a promising platform for tailoring the ground-state properties of materials through the vacuum fluctuations inside the cavity~\cite{flick2017atoms, Ruggi_qed_review_2018, Hannes_chiral_cavity,schlawin2022cavity,lu2025cavity_review,cavity_review_2026}.
The resulting phenomena have been observed in a broad range of experiments, including cavity-altered superconductivity~\cite{Itai_cavity_SC,Kono_cavity_sc}, metal-to-insulator transition~\cite{TaS2_cavity_cdw_nature}, and quantum Hall effect~\cite{appugliese2022breakdown,cavity_QH_prx,Faist_cavity_QHE_2025,Faist_cavity_QHS_NP_2026}. 
On the theoretical side, cavity quantum electrodynamics (cavity QED)-based methods~\cite{ruggenthaler2014_prl_qedft, flick_qedft_oep,pf_QEDFT_pnas,qed_cc_prx_2020,cavity_CIS,cavity_CC_IPEA, cavity_active_space_CI,Demler_cQED_hyperbolic_prl, lr_qed_materials,lu2025cavity_review,QED_HF_enrico_2025} have been developed to describe light-matter interactions in dark cavities, and have been used to predict cavity-induced modifications of 
superconductivity~\cite{sentef_cavity_sc,I-te_mgb2}, polarization~\cite{lin_spontaneous_2026}, magnetization~\cite{emil_cavity_magnetic,Chongxiao_niI2,CIM_effect_Libor} and topology~\cite{Vasil_cavity_topology,cQED_floquet_topo,Dongbin_HgTe_cavity,cavity_graphene_tb,QD_Jiang_cavity_QSL,Hang_qed_hf}.
Despite this progress, most existing studies focus on specific materials and cavity configurations,
and a general principle for determining which macroscopic responses can be induced by cavity vacuum fluctuations---and how these responses depend on the cavity configuration---remains lacking.

\begin{figure}[h]
    \centering
    \includegraphics[width=0.9\linewidth]{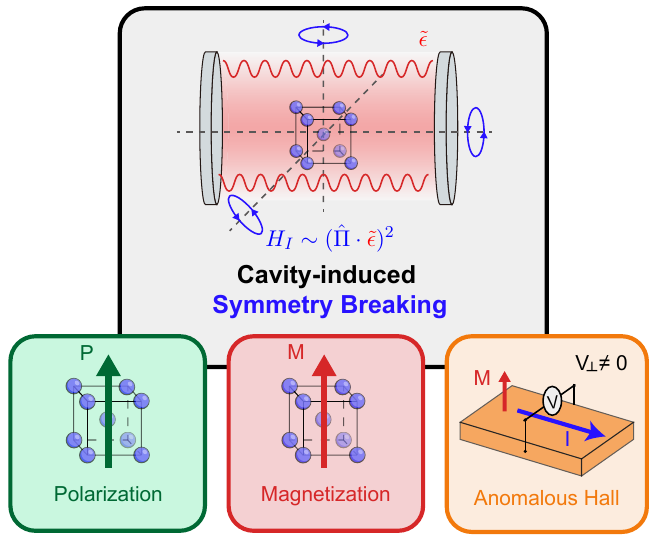}
    \caption{Illustration of cavity-induced symmetry breaking and various resulting phenomena. The blue circular arrows indicate that the symmetry breaking can be tuned continuously by rotating the crystal inside the dark cavity, such that the cavity mode polarization direction $\tilde{\epsilon}$ aligns with different axes.
    $\rm P, M, I$ and $\rm V_{\perp}$ denote the polarization, magnetization, longitudinal current and transverse Hall voltage, respectively.
    }
    \label{fig:Animate_cavity_symmetry_break}
\end{figure}

Symmetry provides a natural framework for addressing this question~\cite{dresselhaus_group_theory,jones_group_theory} and has played an important role in understanding light-matter interactions. 
For instance, crystal symmetry dictates nonlinear-phonon-coupling terms in the transient manipulation of materials with terahertz pulses~\cite{Radaelli_nonlinear_phonon_prb,Cavalleri_nonlinear_phonon_review}. More recently,
dynamical groups combining the spatial, polarization and temporal symmetries of light fields have been developed to explain selection rules in high harmonic generations~\cite{neufeld2019floquet,ofer_light_symmetry_2026}. Floquet optical selection rules combining the symmetries of laser pulses and crystals have also been formulated for pump-probe photoemission spectra~\cite{fan2025floquet,fragkos2026symmetry}. 
In polaritonic chemistry, it has been shown that the symmetry of the cavity can be transmitted to
its vicinity via quantum fluctuations and affect nearby
molecular properties~\cite{QD_Jiang_cavity_chiral,cavity_molecule_symmetry}.

In this Letter, we show that cavity vacuum fluctuations activate free-space-forbidden polarization and magnetization components, even in materials with zero net polarization or magnetization. 
Specifically, we analyze the symmetry of the effective photon-free QED Hamiltonian~\cite{pf_QEDFT_pnas, Pantazopoulos_pf_qed, Hang_qed_hf}, which describes light-matter interaction in a dark cavity. From the symmetry breaking patterns, we identify all crystallographic and magnetic point groups compatible with these effects
and derive the corresponding response tensors relating the induced responses to the cavity modes. 
We then validate these predictions by QEDFT~\cite{pf_QEDFT_pnas,I-te_pxLDA, benshu_qedft} calculations. To illustrate the cavity-induced polarization (CIP) effect, we use $\alpha$-quartz as an example to show how the induced polarization evolves with cavity orientations.
For cavity-induced magnetization (CIM), we use antiferromagnetic (AFM) Mn$_3$Sn as an example, in which the induced magnetization is accompanied by the emergence of an anomalous Hall conductivity (AHC) component that is forbidden outside the cavity.

{\it Symmetry Analysis---}The light-matter interaction in a dark cavity in the high-frequency or strong coupling limit can be described by the effective photon-free QED Hamiltonian, which is derived from the Pauli-Fierz Hamiltonian under dipole approximation~\cite{pf_QEDFT_pnas,I-te_pxLDA,Hang_qed_hf} (in atomic units):
\begin{eqnarray}
\label{eq: photon-free H}
    \hat{H}^{\rm pf} &=& -\frac{1}{2} \sum^{N_e}_{j=1} \nabla^2_j +  \sum^{N_e}_{j=1} v({\bf r}_j) + \frac{1}{2}\sum^{N_e}_{i\neq j} w({\bf r}_i, {\bf r}_j) \nonumber \\
    && + \sum^{M_p}_{\alpha = 1} \frac{\tilde{\omega}_\alpha}{2} - \sum^{M_p}_{\alpha = 1} \frac{\tilde{\lambda}_\alpha^2}{2\tilde{\omega}_\alpha^2} \Big ( \hat{\bf \Pi} \cdot \tilde{{\bm \epsilon}}_\alpha \Big )^2 .
\end{eqnarray}
Here $v({\bf r}_j)$, $w({\bf r}_i, {\bf r}_j)$ and $\hat{\bf \Pi}=-{\mathrm i}\sum^{N_e}_j \nabla_j$ denote the external potential, Coulomb potential and total momentum operator of electrons; 
{$N_e$ and $M_p$ denote the number of electrons in the unit cell and the number of photon modes;} $\tilde\lambda_\alpha$, $\tilde\omega_\alpha$, and $\tilde{\bm \epsilon}_\alpha$ denote the coupling strength, frequency, and polarization direction {of dressed photon mode $\alpha$}. For orthogonal photon modes, these quantities are related to the corresponding bare cavity photon-mode parameters $\lambda_\alpha$, $\omega_\alpha$, and ${\bm \epsilon}_\alpha$  through  $\tilde{\lambda}_\alpha = \lambda_\alpha$, $\tilde{\bm \epsilon}_\alpha = {\bm \epsilon}_\alpha$ and $\tilde{\omega}_{\alpha}^2 = \omega_\alpha^2 + N_e\lambda_\alpha^2$.
We emphasize that “photon-free’’ does not imply the absence of photons in the system. Rather, the influence of the photon modes is effectively encoded in the fluctuation of $\hat{\bf \Pi}$ in the high-frequency or strong coupling limit~\cite{pf_QEDFT_pnas, Simone_sto_pnas, Hang_qed_hf}. 
A more detailed discussion about the photon-free QED Hamiltonian is provided in Sec. I of the Supplementary Material (SM)~\cite{supple}. From Eq.~(\ref{eq: photon-free H}), for each cavity photon mode $\alpha$, the light-matter interaction term is proportional to
\begin{eqnarray}
\label{eq: symmetry of photon-free qed}
     \hat{H}^{\rm pf}_I\sim ({\bf \hat{\bf \Pi}} \cdot \tilde{\bm \epsilon}_\alpha)^2 .
\end{eqnarray}
Therefore, in dark-cavity materials engineering, the symmetry breaking induced by a given cavity configuration can be determined by analyzing how each $({\bf \hat{\Pi}} \cdot \tilde{\bm \epsilon}_\alpha)^2$ transforms under symmetry operations.

From a symmetry perspective, Eq.~(\ref{eq: symmetry of photon-free qed}) preserves inversion ($\mathcal P$) and time-reversal ($\mathcal{T}$) symmetries, since both operations transform ${\bf \Pi}\to -{\bf \Pi}$, leaving the squared coupling invariant.
This contrasts with a classical electromagnetic field, which couples linearly to $\hat{\bf \Pi}$ and can therefore break $\mathcal P$ and $\mathcal T$ symmetries. Following the same reasoning, we analyze all crystal symmetry operations and determine whether they are preserved or broken by linearly polarized cavity modes.
One key consequence is that a cavity mode $\alpha$ breaks rotational symmetries $\mathcal{C}_n$ ($n=3,4,6$) whose rotation axes are not parallel to $\tilde{\bm \epsilon}_\alpha$, and $\mathcal{C}_2$ whose rotation axes that are neither parallel nor perpendicular to $\tilde{\bm \epsilon}_\alpha$, since these rotations change the projection of $\bf \Pi$ onto $\tilde{\bm \epsilon}_\alpha$. As a result, the associated composite symmetries $\mathcal{C}_n \mathcal{P}$ and $\mathcal{C}_n \mathcal{T}$ are also broken.
Such symmetry breaking can lead to a broad range of material modifications, as illustrated in Fig.~\ref{fig:Animate_cavity_symmetry_break}.

Theoretically, zero macroscopic polarization (magnetization) can be protected by $\mathcal{P}$ ($\mathcal{T}$) symmetry, or suitable combinations of  $\mathcal{C}_n$, $\mathcal{C}_n \mathcal{P}$, and $\mathcal{C}_n \mathcal{T}$ ($n=2,3,4,6$) symmetries~\cite{dresselhaus_group_theory,bilbao_paper}.
Based on the analysis above, although $\mathcal{P}$, $\mathcal{T}$ symmetries cannot be broken by cavity modes considered here, one can still choose an appropriate $\tilde{\bm \epsilon}_\alpha$ to break the relevant $\mathcal{C}_n$, $\mathcal{C}_n \mathcal{P}$ or $\mathcal{C}_n \mathcal{T}$ symmetries and induce a finite macroscopic polarization or magnetization. 
We first consider the point groups (PGs) that can host the CIP effect. Such PGs must not contain $\mathcal{P}$, yet can still contain several $\mathcal{C}_n$ or $\mathcal{C}_n\mathcal{P}$ operations that enforce zero net polarization. Applying the symmetry-breaking rules of cavity modes discussed above, we identify ten nonpolar PGs compatible with the CIP effect, as listed in Tab.~\ref{tab: CIP CIM groups}. Similarly, for the CIM effect, the relevant magnetic point groups (MPGs) must not contain $\mathcal{T}$ or $\mathcal{PT}$, but could contain $\mathcal{C}_n$, $\mathcal{C}_n\mathcal{P}$ or $\mathcal{C}_n\mathcal{T}$ that protect zero magnetization. In Tab.~\ref{tab: CIP CIM groups}, we also list the non-ferromagnetic MPGs compatible with the CIM effect. 
More generally, these effects can also occur in materials with intrinsic polarization (magnetization). The full (M)PG tables including these cases are provided in Sec. II of the SM~\cite{supple}.

\begin{table}[h]
    \centering
    \begin{tabular}{ cc
    @{\hspace{0.5cm}} 
    cccc}
    \toprule
        \multicolumn{2}{c@{\hspace{0.5cm}}}{CIP (PG)}  & \multicolumn{4}{c}{CIM (MPG)}  \\
    \midrule
         D$_2$ & D$_{2d}$ 
          & $222.1$ & $mm2.1$ & $mmm.1$ & $4'$ \\
        D$_3$ & C$_{3h}$ 
        & $\bar4'$ &  $4'/m$ & $422.1$ & $4'22'$ \\
        D$_{3h}$ & S$_4$ 
        & $4mm.1$ & $4'm'm$ & $\bar42m.1$ & $\bar4'2'm$ \\
        D$_4$ & D$_6$ 
        & $\bar4'2m'$ & $4/mmm.1$ & $4'/mm'm$ & $32.1$ \\
        T  & T$_d$ 
        & $3m.1$ & $\bar3m.1$ & $6'$ & $\bar6'$   \\
        &  & $6'/m'$ & $622.1$ & $6'22'$ & $6mm.1$  \\
        &  & $6'mm'$ & $\bar6m2.1$ & $\bar6'm'2$ & $\bar6'm2'$  \\
        &  & $6/mmm.1$ & $6'/m'mm'$ & $23.1$ & $m\bar3.1$  \\
        &  & $4'32'$ & $\bar4'3m'$ & $m\bar3m'$ \\
    \bottomrule
    \end{tabular}
    \caption{All nonpolar point groups (PGs) compatible with the CIP effect, and all non-ferromagnetic magnetic point groups (MPGs) that are compatible with the CIM effect.}
    \label{tab: CIP CIM groups}
\end{table}

Beyond identifying the relevant (M)PGs, symmetry analysis also allows us to establish a relation between the light-matter coupling strength, the cavity mode polarization directions, 
and the induced polarization.
For this purpose, we formulate the CIP effect in terms of response tensors with respect to the cavity modes:
$P_i\sim d_{ij}(\tilde\lambda/\tilde\omega) \tilde{\epsilon}_j + d_{ijk}(\tilde\lambda/\tilde\omega)^2\tilde{\epsilon}_j\tilde{\epsilon}_k+ \mathcal{O}((\tilde\lambda/\tilde\omega)^3)$,
where $d_{ij}$ and $d_{ijk}$ are different order response tensors. This formulation has the advantage of providing a unified description of the induced polarization for general cavity environments. According to Eq.~(\ref{eq: symmetry of photon-free qed}), the effective light-matter coupling in a dark cavity depends quadratically on the dressed cavity fields and preserves $\mathcal{P}$ symmetry. As a result, all odd-order terms in $\tilde{\bm \epsilon}$ vanish,
and the leading-order CIP response is: 
\begin{eqnarray}
\label{eq: CIP response formula}
    {P}_i = \sum_\alpha \left(\frac{\tilde{\lambda}_\alpha}{\tilde{\omega}_\alpha}\right)^2
    d_{ijk} \tilde\epsilon_{j,\alpha} \tilde\epsilon_{k,\alpha} \equiv\sum_\alpha\left(\frac{\tilde{\lambda}_\alpha}{\tilde{\omega}_\alpha}\right)^2{d}_{l(i)m} \bar{\epsilon}_{m,\alpha}.\nonumber\\
\end{eqnarray}
Here, $P_i$ denotes the polarization in $i\in\{x,y,z\}$, and $d_{ijk}$ denotes the third-rank CIP tensor, which is symmetric under permutation $j \leftrightarrow k$. 
The quantities $d_{lm}$ and $\bar{\epsilon}_m$ are the corresponding components in Voigt notation (see Sec. III of the SM~\cite{supple} for the convention) and $\alpha$ runs over all cavity modes.
The material's PG symmetry further constrains the tensor elements $d_{lm}$~\cite{dresselhaus_group_theory,jones_group_theory}. In Sec. III of the SM~\cite{supple}, we also discuss how to derive the independent $d_{lm}$ elements and provide the CIP tensors for all PGs listed in Tab.~\ref{tab: CIP CIM groups}. 
Similarly, the CIM effect can also be described in terms of a response tensor as
\begin{eqnarray}
       M_i &=& \sum_\alpha \left(\frac{\tilde{\lambda}_\alpha}{\tilde{\omega}_\alpha}\right)^2\Lambda_{l(i)m} \bar\epsilon_{m,\alpha} ,
       \label{eq: CIM tensor form}
\end{eqnarray}
where $M_i$ is the induced magnetization in direction $i$, and $\Lambda$ is the CIM tensor in Voigt notation.

\begin{center}
    \begin{figure*}[ht!]
        \centering
        \includegraphics[width=1\linewidth]{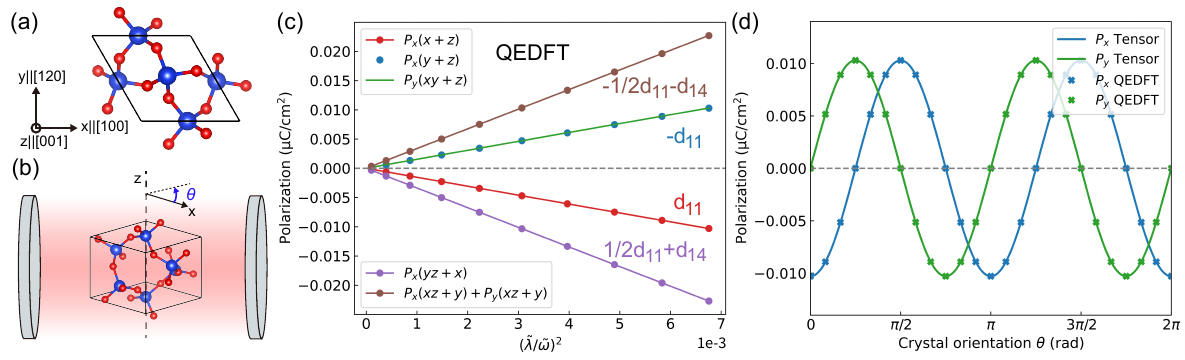}
        \caption{(a) Illustration of the crystal structure of $\alpha$-quartz.
        (b) Illustration of rotating the Fabry-P\'{e}rot cavity with the crystal fixed inside, such that the cavity mode polarization plane aligns with different crystal planes.
        (c) Polarization of $\alpha$-quartz for different cavity configurations (crystal orientations) calculated by QEDFT. The slopes of the linear curves correspond to the respective CIP tensor elements.
        (d) Polarization of $\alpha$-quartz as a function of cavity rotation angle $\theta$ in the $xy$ plane. The results from QEDFT and Eq.~(\ref{eq: P angle from symmetry}) are in full agreement.
        }
        \label{fig:quartz_polar}
    \end{figure*}
\end{center}

{\it Cavity-induced Polarization---}In this section, we illustrate the CIP effect using $\alpha$-quartz (SiO$_2$), a material widely used in optical and electric applications~\cite{quartz_review}, as a concrete example. 
The computational details are summarized in Sec. IV of the SM~\cite{supple}.
$\alpha$-quartz is a 
chiral crystal belonging to space group $P3_221$ (or $P3_121$), with point group D$_3$. Here we choose the $P3_221$ enantiomorph with the crystal structure shown in Fig.~\ref{fig:quartz_polar}(a). 
The corresponding CIP tensor for D$_3$ group is~\cite{supple}:
\begin{eqnarray}
d =   
    \begin{pmatrix}
        d_{11} & -d_{11} & 0 & d_{14} & 0 & 0 \\
        0 & 0 & 0 & 0 & -d_{14} & -d_{11} \\
        0 & 0 & 0 & 0 & 0 & 0 
    \end{pmatrix} .
    \label{eq: CIP SiO2}
\end{eqnarray}
The coordinate system used to define this tensor is shown in Fig.~\ref{fig:quartz_polar}(a). The cavity-mode directions introduced below are expressed in the same coordinate system. In a Fabry-P\'{e}rot cavity, the cavity field can be effectively described by two orthogonal modes polarized in the plane parallel to the cavity mirrors~\cite{lu2025cavity_review}. We therefore focus here on two orthogonal cavity modes with identical $\lambda$ and $\omega$.
The analysis can, however, be straightforwardly extended to other cavity quantum environments.
In this setup, different $d_{lm}$ elements can be accessed by rotating the cavity (or the crystal inside the cavity) so that the cavity polarization plane aligns with different crystal planes. 
According to Eqs.~(\ref{eq: CIP response formula}) and (\ref{eq: CIP SiO2}), two orthogonal modes polarized along $x$ ($\tilde{\bm \epsilon} = (1,0,0)$) $+\ z$ ($\tilde{\bm \epsilon} = (0,0,1)$), or along $y$ ($\tilde{\bm \epsilon} = (0,1,0)$) $+\ z$, with identical coupling strength and frequency, should induce polarizations of opposite sign in $x$:
\begin{eqnarray}
    P_x(x+z ) = \frac{\tilde{\lambda}^2}{\tilde{\omega}^2} d_{11}, \
    P_x(y+z) = -\frac{\tilde{\lambda}^2}{\tilde{\omega}^2} d_{11}.
\end{eqnarray}
Here the notation $P_i(\sum_\alpha\tilde{\bm\epsilon}_\alpha)$ denotes the induced polarization component $i$ by a set of cavity photon modes $\alpha$. 
This prediction is confirmed by our QEDFT calculations, as shown in Fig.~\ref{fig:quartz_polar}(c). Moreover, other components of the CIP tensor can be accessed by choosing cavity modes polarized along off-diagonal directions ($xy$, $xz$ and $yz$). For a cavity mode polarized along $xy$ ($\tilde{\bm \epsilon} = (1/\sqrt{2},1/\sqrt{2},0)$), the relevant Voigt components (i.e., nonzero $\tilde{\bm \epsilon}\tilde{\bm \epsilon}$ elements) are: $ 1/2\bar{\epsilon}_1, {1}/{2}\bar{\epsilon}_2$, and $\bar{\epsilon}_6 $.
Therefore, for two orthogonal cavity modes in the $xy+z$ configuration, the induced polarization should be
$P_x(xy+z) = 0$ and $P_y(xy+z) = -\frac{\tilde{\lambda}^2}{\tilde{\omega}^2} d_{11}$,
in agreement with the QEDFT results in Fig.~\ref{fig:quartz_polar}(c).
Similarly, the Voigt elements associated with $xz$ ($\tilde{\bm \epsilon} = (1/\sqrt{2},0,1/\sqrt{2})$) + $y$ and $yz$ ($\tilde{\bm \epsilon} = (0,1/\sqrt{2},1/\sqrt{2})$) + $x$ cavity modes are $1/2(\bar{\epsilon}_1+ \bar{\epsilon}_3 )+ \bar\epsilon_5+ \bar{\epsilon}_2$ and $1/2(\bar{\epsilon}_2+\bar{\epsilon}_3 )+\bar{\epsilon}_4+\bar{\epsilon}_1$, respectively. Therefore,
the induced polarizations for these two cavity mode configurations satisfy:
\begin{eqnarray}
    %P_x( \tilde{\bm \epsilon} = (0,1,1) ) = - P_x( \tilde{\bm \epsilon} = (1,0,1) ) - P_y( \tilde{\bm \epsilon} = (1,0,1) ). \nonumber
    \Big|P_x(xz+y)+P_y(xz+y)\Big|=\Big|P_x(yz+x)\Big|,
\end{eqnarray}
which is also confirmed by the QEDFT results in Fig.~\ref{fig:quartz_polar}(c).

To demonstrate that cavity can serve as a continuous knob for materials engineering, we calculate the evolution of the induced polarization as the cavity is rotated. We consider a cavity configuration with two modes originally polarized along $x$ and $z$. Then, as illustrated in Fig.~\ref{fig:quartz_polar}(b), we rotate the cavity (or equivalently, rotate the $\alpha$-quartz crystal) about the $z$ axis
% in the $xy$ plane 
and evaluate the induced polarization as a function of the rotation angle $\theta$ to $x$ in the $xy$-plane. From the CIP tensor in Eq.~(\ref{eq: CIP SiO2}), the induced polarization should evolve as
\begin{equation}
\begin{aligned}
\label{eq: P angle from symmetry}
P_x(\theta) &= \frac{\tilde{\lambda}^2}{\tilde{\omega}^2}d_{11} (\cos^2\theta - \sin^2\theta), \\
P_y(\theta) &= -2\frac{\tilde{\lambda}^2}{\tilde{\omega}^2}d_{11} \cos\theta\sin\theta,
\end{aligned}
\end{equation}
where $d_{11}$ is fitted from the QEDFT results of $P_x(x+z)$  in Fig.~\ref{fig:quartz_polar}(c).
As shown in Fig.~\ref{fig:quartz_polar}(d), the QEDFT results for $\tilde{\lambda}^2/\tilde{\omega}^2=0.007$ with varying cavity orientation are in excellent agreement with the symmetry-based predictions.

Besides $\alpha$-quartz, we also perform extensive first principles QEDFT calculations to verify the CIP effect for all ten predicted CIP-compatible nonpolar PGs, using one representative material for each group, as presented in Sec. V of the SM~\cite{supple}.

\begin{figure}[ht!]
        \centering
        \includegraphics[width=0.8\linewidth]{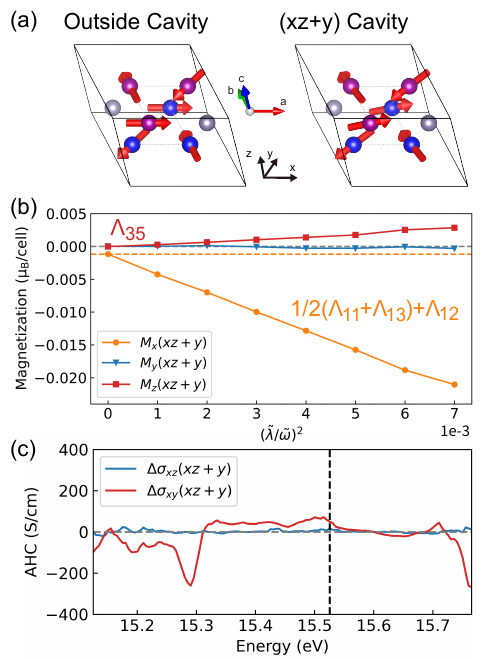}
        \caption{(a) Crystal structure of AFM-1 Mn$_3$Sn with local magnetic moments.
        Left: outside the cavity. Right: coupled to $xz+y$ cavity modes.
        Purple and blue spheres denote Mn atoms in different layers.
        (b) Magnetization under $xz+y$ cavity modes obtained from QEDFT.
        The slopes of the linear curves relate to the corresponding CIM tensor elements.
        (c) Changes in the AHC components  in and outside cavity ($\Delta\sigma$) as a function of energy relative to the Fermi level. The black dashed line indicates the Fermi level. 
        }
        \label{fig:Mn3Sn_mag}
    \end{figure}

{\it Cavity-induced Magnetization---}In this section, we take the widely studied coplanar AFM Mn$_3$Sn in the AFM-1 phase~\cite{zhangYang_mn3sn_prb, mn3sn_piezomagnetic_prb,mn3sn_mx_exp_NM,mn3sn_mx_exp_NP} as a representative example to demonstrate the CIM effect and the corresponding response tensor in Eq.~(\ref{eq: CIM tensor form}).
As shown in Fig.~\ref{fig:Mn3Sn_mag}(a), the crystal structure of AFM-1 Mn$_3$Sn consists of two layers of staggered Mn triangles forming a kagome lattice, belonging to space group $P6_3/mmc$ ($\#194$) with point group $\mathrm{D}_{6h}$. When magnetic order and spin-orbit coupling (SOC) are taken into account, its symmetry is described by MPG $mm'm'$ ({\#8.4.27})~\cite{mn3sn_piezomagnetic_prb}, 
which allows a weak net magnetization along $x$. In our DFT calculations without a cavity, we obtain $M^0_x \sim 0.0015 {\rm \mu_B}/{\rm cell}$, in good agreement with experimental observations~\cite{mn3sn_mx_exp_NP, mn3sn_mx_exp_NM}. 
The corresponding CIM tensor for $mm'm'$ is (see Sec. VI of the SM~\cite{supple}):
\begin{eqnarray}
        \Lambda = \begin{pmatrix}
        \Lambda_{11} & \Lambda_{12} & \Lambda_{13} & 0 & 0 & 0 \\
        0 & 0 & 0 & 0 & 0 & \Lambda_{26} \\
        0 & 0 & 0 & 0 & \Lambda_{35} & 0 
    \end{pmatrix}.
    \label{eq: Mn3Sn CIM tensor}
\end{eqnarray}
Fig.~\ref{fig:Mn3Sn_mag}(b) shows the induced magnetization obtained from QEDFT for two orthogonal cavity modes in the ($xz+y$) configuration. Consistent with Eq.~(\ref{eq: Mn3Sn CIM tensor}), a finite $M_z$ is induced by the cavity modes in addition to $M_x$:
\begin{equation}
\begin{aligned}
% \begin{eqnarray}
    M_x(xz+y) &=  M^0_x+\frac{\tilde{\lambda}^2}{\tilde{\omega}^2}[1/2(\Lambda_{11}+\Lambda_{13})+\Lambda_{12}], 
    % \nonumber 
    \\ M_z(xz+y) &= \frac{\tilde{\lambda}^2}{\tilde{\omega}^2}\Lambda_{35},
% \end{eqnarray}
\end{aligned}
\end{equation}
while $M_y$ remains zero. The right panel of Fig.~\ref{fig:Mn3Sn_mag}(a) illustrates the corresponding change in the local magnetic moments: two in-plane spins cant out of plane, producing a net magnetization along $z$.
The symmetry breaking induced by this cavity mode configuration also enables an AHC component that is forbidden outside the cavity. In pristine AFM-1 Mn$_3$Sn, for the Berry curvature $\mathbf \Omega=(\Omega_{yz},\Omega_{xz},\Omega_{xy})$, the mirror symmetry operations $\mathcal{M}_x$ and 
$\mathcal{M}_y\mathcal{T}$ transform it as $\mathcal{M}_x\mathbf{\Omega}=(\Omega_{yz},-\Omega_{xz},-\Omega_{xy})$ and $(\mathcal{M}_y\mathcal{T})\mathbf{\Omega}=(\Omega_{yz},-\Omega_{xz},\Omega_{xy})$, respectively. Since the AHC is obtained by integrating the Berry curvature over the Brillouin zone, 
$\sigma_{yz}$ is the only symmetry-allowed AHC component, whereas $\sigma_{xz}$ and $\sigma_{xy}$ must vanish. When the CIM is along the $z$ direction, $\mathcal{M}_x$ is broken, thereby lifting the symmetry constraint on $\sigma_{xy}$. However, the combined symmetry $\mathcal{M}_y\mathcal{T}$ remains intact, which still enforces $\sigma_{xz}=0$. Therefore, the ($xz+y$) cavity mode configuration induces a finite $\sigma_{xy}$, while $\sigma_{xz}$ remains zero, in agreement with our QEDFT results. As shown in Fig.~\ref{fig:Mn3Sn_mag}(c), at $\tilde{\lambda}^2/\tilde{\omega}^2=0.007$, the cavity induces $\sigma_{xy}\sim 49$ S/cm at the Fermi level, compared with the intrinsic AHC component $\sigma_{yz}\sim 184$ S/cm. In Sec. VII of the SM~\cite{supple}, we estimate how the dressed coupling strength relates to possible experimental setups.

Finally, our symmetry analysis can be naturally extended to spin groups~\cite{libor_prx_altermagnet_2022,qihang_prx_spin_group_2022} when SOC is negligible. In Sec. VIII of the SM~\cite{supple}, we derive the CIM tensor for the spin point group of AFM-1 Mn$_3$Sn and find full consistency with QEDFT calculations without SOC.
Note that in the absence of SOC, spin and lattice rotations are decoupled, 
and $M_z$ is therefore forbidden for all cavity configurations by the spin-space rotational symmetry. When SOC is included, spin-only rotations are no longer valid symmetries, allowing $M_z$ to be induced.

{\it Conclusion---}In summary, guided by symmetry analysis of the effective photon-free QED Hamiltonian,
we demonstrate that cavity vacuum fluctuations can induce macroscopic polarization and magnetization components that are strictly forbidden in free space. We identify all (M)PGs compatible with such responses and derive the corresponding leading-order cavity-response tensors. Our QEDFT calculations reveal that the induced polarization in $\alpha$-quartz can be continuously and precisely tuned by rotating the cavity orientation. In AFM-1 Mn$_3$Sn, cavity-induced symmetry breaking generates an out-of-plane magnetization component and unlocks an AHC component that is entirely absent without the cavity.
Together, these results establish a general symmetry-based framework for predicting and controlling cavity-modified polarization and magnetization in quantum materials. Looking ahead, this framework opens new avenues for exploring cavity-enabled functionalities such as multiferroicity, chirality, and altermagnetism. A particularly promising direction is to go beyond the dipole approximation by incorporating the spatial structure of cavity modes, which can break inversion and other symmetries, unlocking a broader class of cavity-induced phenomena.

{\it Acknowledgments---}We would like to thank Zhiyang Zeng, Xinle Cheng, Michael Ruggenthaler, and Hannes H{\"u}bener for fruitful discussions. We thank Professor Peizhe Tang for his critical reading and valuable feedback on the manuscript. J.Q. acknowledges support from
the Max-Planck Graduate Center for Quantum materials. This work was supported by the European Research Council (ERC-2024-SyG-101167294; UnMySt) and the Cluster of Excellence Advanced Imaging of Matter (AIM). We acknowledge support from the Max Planck-New York City Center for Non-Equilibrium Quantum Phenomena. The Flatiron Institute is a division of the Simons Foundation.

\nocite{
bilbao_paper,
bilbao_magnetic_group_tensor,
bilbao_spin_group_tensor,
qihang_prx_spin_group_2024,
qe_2020,
ONCV_pseudopotential,
pseudodojo,
pbe,
king_vanderbilt_modern_polarization,
vanderbilt_berry_phase_book,
Wang_ahc_wannier,
wannier90_jpcm,
materials_project,
litvin_spin_point_group,
Mark_effective_single_mode_2025}

\bibliography{main}% Produces the bibliography via BibTeX.

%\nocite{*}

\end{document}

\preprint{APS/123-QED}

\title{Supplementary Material for \\
``Macroscopic Polarization and Magnetization from Cavity Vacuum Fluctuations''
}% Force line breaks with \\

\author{Jingkai Quan}
\email{jingkai.quan@mpsd.mpg.de}
\affiliation{%
Max Planck Institute for the Structure and Dynamics of Matter,
Luruper Chausse 149, 22761, Hamburg, Germany
}%
\author{Chongxiao Fan}%
\email{chongxiao.fan@mpsd.mpg.de}
\affiliation{%
Max Planck Institute for the Structure and Dynamics of Matter,
Luruper Chaussee 149, 22761 Hamburg, Germany
}
\affiliation{Institute for Theory of Statistical Physics, RWTH Aachen University, and JARA Fundamentals of Future Information Technology, 52062 Aachen, Germany}
\author{Benshu Fan}
\email{benshu.fan@mpsd.mpg.de}
\affiliation{%
Max Planck Institute for the Structure and Dynamics of Matter,
Luruper Chaussee 149, 22761 Hamburg, Germany
}%
\author{I-Te Lu}
\affiliation{%
Max Planck Institute for the Structure and Dynamics of Matter,
Luruper Chaussee 149, 22761 Hamburg, Germany
}%
\author{Dante M. Kennes}
\affiliation{Institute for Theory of Statistical Physics, RWTH Aachen University, and JARA Fundamentals of Future Information Technology, 52062 Aachen, Germany}
\affiliation{%
Max Planck Institute for the Structure and Dynamics of Matter,
Luruper Chaussee 149, 22761 Hamburg, Germany
}%
\author{Angel Rubio}
 \email{angel.rubio@mpsd.mpg.de}
\affiliation{%
Max Planck Institute for the Structure and Dynamics of Matter,
Luruper Chaussee 149, 22761 Hamburg, Germany
}%
% \affiliation{
% Initiative for Computational Catalysis (ICC), The Flatiron Institute, 162 Fifth Avenue, New York, NY 10010, United States
% }
\affiliation{Initiative for Computational Catalysis, The Flatiron Institute, Simons Foundation, New York City, NY 10010, United States of America}

\date{\today}% It is always \today, today,

\maketitle

%\tableofcontents

\setcounter{equation}{0}
\renewcommand{\theequation}{S\arabic{equation}}

\setcounter{figure}{0}
\renewcommand{\thefigure}{S\arabic{figure}}

\setcounter{table}{0}
\renewcommand{\thetable}{S\arabic{table}}

\section{The Effective Photon-free QED Hamiltonian}

Within the long-wavelength approximation and the Coulomb gauge, the Pauli-Fierz (PF) Hamiltonian of non-relativistic quantum electrodynamics (QED)~\cite{pf_QEDFT_pnas,lu2025cavity_review} is written as
\begin{eqnarray}
\label{eq: PF original}
    \hat{H}_{\rm PF} = \frac{1}{2} \sum^{N_e}_{j=1} \Big ( -{\mathrm i}\nabla_j + \frac{1}{c} \hat{\bf A} \Big)^2 
    + \sum^{N_e}_{j=1} v({\bf r}_j)
    + \frac{1}{2}\sum^{N_e}_{i\neq j} w({\bf r}_i, {\bf r}_j) + \sum^{M_p}_{\alpha = 1} \omega_\alpha \Big ( \hat{a}^\dagger_\alpha \hat{a}_\alpha + \frac{1}{2} \Big ) .
\end{eqnarray}
For the matter sector, $v({\bf r}_j)$, $w({\bf r}_i, {\bf r}_j)$, and $N_e$ denote the external potential, the electron-electron Coulomb interaction, and the number of electrons in the unit cell. For the photonic sector,
$\hat{\bf A}$, $\hat{a}$, $\hat{a}^\dagger$, $\omega$, and $M_p$ denote the vector potential, photon annihilation, photon creation operator, photon mode  frequency and the number of photon modes, respectively.
The vector potential $\hat{\bf A}$  can be expressed as
\begin{eqnarray}
    \hat{\bf A} = c \sum^{M_p}_{\alpha = 1} \lambda_\alpha {\bm \epsilon}_\alpha \frac{1}{\sqrt{2\omega_\alpha}} (\hat{a}^\dagger_\alpha + \hat{a}_\alpha) .
    \label{eq: vector potential}
\end{eqnarray}
Here, $\lambda_\alpha$ and ${\bm \epsilon}_\alpha$ denote the coupling strength and polarization direction of photon mode $\alpha$.
Eq.~(\ref{eq: PF original}) can be simplified by absorbing 
the diamagnetic term $\hat{\bf A}^2$ into the photon operators by a Bogoliubov transformation~\cite{pf_QEDFT_pnas,I-te_pxLDA}. The resulting Hamiltonian can be written in terms of dressed photon modes
\begin{eqnarray}
    \hat{\tilde{H}}_{\rm PF} = -\frac{1}{2} \sum^{N_e}_{j=1} \nabla^2_j +  \sum^{N_e}_{j=1} v({\bf r}_j) + \frac{1}{2}\sum^{N_e}_{i\neq j} w({\bf r}_i, {\bf r}_j) 
    + \frac{1}{c} \hat{\tilde{ {\bf A} }} \cdot \hat{ {\bf \Pi} } + \sum^{M_p}_{\alpha = 1} \tilde{\omega}_\alpha \Big ( \hat{\tilde{a}}^\dagger_\alpha \hat{\tilde{a}}_\alpha + \frac{1}{2} \Big ) ,
    \label{eq: HPF dressed}
\end{eqnarray}
where $\hat{\bf \Pi}= -{\rm i}\sum_j^{N_e} \nabla_j$, and the dressed vector potential $\hat{\tilde{\bf A}}$ is defined as
\begin{eqnarray}
    \hat{\tilde{ {\bf A}} } = c \sum_\alpha \tilde\lambda_\alpha \tilde{\bm \epsilon}_\alpha \frac{1}{\sqrt{2 \tilde{\omega}_\alpha}} (\hat{\tilde{a}}^\dagger_\alpha + \hat{\tilde{a}}_\alpha) .
    \label{eq: vector potential bogo}
\end{eqnarray}
For mutually orthogonal photon modes, the Bogoliubov transformation simply yields $\tilde{\lambda}_\alpha = \lambda_\alpha$, $\tilde{\bm \epsilon}_\alpha = {\bm \epsilon}_\alpha$, and $\tilde{\omega}_{\alpha}^2 = \omega_\alpha^2 + N_e\lambda_\alpha^2$. The Hamiltonian~(\ref{eq: HPF dressed}) acts in the hybrid light-matter Hilbert space. Its eigenstates should therefore be expanded in a basis formed by the electronic eigenstates and the Fock states of the dressed photon modes, $\{\ket{\tilde0},\ket{\tilde1},\ket{\tilde2}, \cdots\}$. 
However, in the high-frequency or strong coupling limit, different photon sectors become off-resonant. As a result, the PF Hamiltonian can then be downfolded onto the zero dressed-photon sector~\cite{pf_QEDFT_pnas,Simone_sto_pnas,Hang_qed_hf}.
In other words, in this limit, the quantum fluctuations of the vector potential $\hat{\tilde{\bf A}}$ can be approximated in terms of $\hat{\bf \Pi}$~\cite{pf_QEDFT_pnas}. This procedure effectively absorbs the photonic degrees of freedom into the electronic degrees of freedom. Under this approximation, the PF Hamiltonian of a time-independent system reduces to
\begin{eqnarray}
\label{eq: photon-free H}
    \hat{H}^{\rm pf} = -\frac{1}{2} \sum^{N_e}_{i=1} \nabla^2_i +  \sum^{N_e}_{i=1} v({\bf r}_i) + \frac{1}{2}\sum^{N_e}_{i\neq j} w({\bf r}_i, {\bf r}_j)
     + \sum^{M_p}_{\alpha = 1} \frac{\tilde{\omega}_\alpha}{2} - \sum^{M_p}_{\alpha = 1} \frac{\tilde{\lambda}_\alpha^2}{2\tilde{\omega}_\alpha^2} \Big ( \hat{\bf \Pi} \cdot \tilde{{\bm \epsilon}}_\alpha \Big )^2 ,
\end{eqnarray}
which is referred to as the effective photon-free QED Hamiltonian~\cite{pf_QEDFT_pnas}. Within the framework of quantum electrodynamical density functional theory (QEDFT), one can construct an auxiliary non-interacting system that reproduces the ground-state electronic density of Eq.~(\ref{eq: photon-free H}). For instance, the electron-photon exchange local density approximation (pxLDA) functional is given by~\cite{I-te_pxLDA,benshu_qedft}:
\begin{eqnarray}
\label{eq: poisson pxlda}
    \nabla^2 v_{\rm pxLDA}({\bf r}) = - \sum_\alpha \frac{2\pi^2 \tilde{\lambda}^2_\alpha}{\tilde{\omega}^2_\alpha}
    \left \{ (\tilde{\bm \epsilon}_\alpha \cdot \nabla)^2 \left [  \frac{\rho({\bf r})}{2V}\right]^{\frac{2}{3}}\right\},
\end{eqnarray}
where $V$ and $\rho({\bf r})$ are the unit cell volume and the electron density.

\section{Full Point group list of the CIP and CIM effect}
In the main text, we only present the nonpolar point groups and non-ferromagnetic magnetic point groups that are compatible with  cavity-induced polarization (CIP) and cavity-induced magnetization (CIM) effects. However, these effects can also exist in polar/ferromagnetic groups, i.e., cavity can also modify polarization and magnetization in pristine ferroelectric or magnetic materials. Here in Tabs.~\ref{tab: cip full pg} and \ref{tab: cim full pg}, we give the full list of the point groups (PGs) and magnetic point groups (MPGs) that are compatible with these effects, respectively.

\begin{table}[htbp]
\centering
\begin{tabular}{ *{5}c@{\hspace{0.7cm}}*{5}c}
\toprule
 \multicolumn{10}{c}{CIP PGs} \\
  \multicolumn{5}{c}{nonpolar} & \multicolumn{5}{c}{polar}\\
  \midrule
  D$_2$ & D$_{2d}$ & D$_3$ & C$_{3h}$ & D$_{3h}$ &
  C$_1$ & C$_2$ & C$_{s}$ & C$_{2v}$ & C$_4$
  \\
  S$_4$ & D$_4$ & D$_6$ & T  & T$_d$ & 
  C$_{4v}$ & C$_3$ & C$_{3v}$ & C$_{6}$ & C$_{6v}$
  \\
  \bottomrule
    \end{tabular}
    \caption{List of all CIP-compatible point groups.}
    \label{tab: cip full pg}
\end{table}

\begin{table}[htbp]
\centering
\begin{tabular}{ *{5}c@{\hspace{0.7cm}}*{5}c}
\toprule
 \multicolumn{10}{c}{CIM MPGs} \\
  \multicolumn{5}{c}{non-ferromagnetic} & \multicolumn{5}{c}{ferromagnetic}\\
  \midrule
   $222.1$ & $mm2.1$ & $mmm.1$ & $4'$ & $\bar4'$ &
  $1.1$ & $\bar1.1$ & $2.1$ &  $2'$ &  $m.1$ 
  \\
  $4'/m$ & $422.1$ & $4'22'$ & $4mm.1$ & $4'm'm$ & 
  $m'$ & $2/m.1$ & $2'/m'$ & $2'2'2$ & $m'm2'$
  \\
  $\bar42m.1$ & $\bar4'2'm$ & $\bar4'2m'$ & $4/mmm.1$ & $4'/mm'm$ &
  $m'm'2$ & $mm'm'$ & $4.1$ & $\bar4.1$ & $4/m.1$
  \\
   $32.1$ & $3m.1$ & $\bar3m.1$ & $6'$ & $\bar6'$ &
   $42'2'$ & $4m'm'$ & $\bar42'm'$ & $4/mm'm'$ & $3.1$
   \\
    $6'/m'$ & $622.1$ & $6'22'$ & $6mm.1$ & $6'mm'$ & 
   $\bar3.1$ & $32'$ & $3m'$ & $\bar3m'$ & $6.1$
   \\
    $\bar6m2.1$ & $\bar6'm'2$ & $\bar6'm2'$ & $6/mmm.1$ & $6'/m'mm'$ &
   $\bar6.1$ & $6/m.1$ & $62'2'$ & $6mm'$ & $\bar6m'2'$
   \\
     $23.1$ & $m\bar3.1$ & $4'32'$ & $\bar4'3m'$ & $m\bar3m'$ &
   $6/mm'm'$ 
   \\    
  \bottomrule
    \end{tabular}
    \caption{List of all CIM-compatible magnetic point groups.}
    \label{tab: cim full pg}
\end{table}

\section{Derivation of the independent CIP tensor elements}
\label{app: cip tensor}

In Voigt notation, the CIP tensor can be written as:
\begin{eqnarray}
    {\mathbf P}=\begin{pmatrix}
        P_x \\
        P_y \\
        P_z 
    \end{pmatrix} =   \begin{pmatrix}
        d_{11} & d_{12} & d_{13} & d_{14} & d_{15} & d_{16} \\
        d_{21} & d_{22} & d_{23} & d_{24} & d_{25} & d_{26} \\
        d_{31} & d_{32} & d_{33} & d_{34} & d_{35} & d_{36} 
    \end{pmatrix} 
    \begin{pmatrix}
        \bar\epsilon_{1} \\
        \bar\epsilon_{2} \\
        \bar\epsilon_{3} \\
        \bar\epsilon_{4} \\
        \bar\epsilon_{5} \\
        \bar\epsilon_{6} 
    \end{pmatrix},
    \label{eq: CIP general}
\end{eqnarray}
where the correspondence between the Voigt components and the standard tensor elements is given in Tab.~\ref{tab: voigt}. Here, we adopt the convention that the factor of 2 is included in the Voigt representation of the shear photon polarization components, e.g. $\bar\epsilon_6=2\tilde{\epsilon}_x\tilde{\epsilon}_y$. Alternatively, one may choose to absorb this factor of 2 into the Voigt CIP tensor elements, e.g. $d'_{16}=2d_{xyz}$, which leads to a different but equivalent tensor. Both Voigt conventions are common in the literature.
We have also neglected the coupling strength ${\tilde{\lambda}}^2/ {\tilde{\omega}}^2$ for simplicity, which will not affect our derivation.
The induced polarization should change as coordinate $(x,y,z)$ under all symmetry operations of the corresponding point group, while the CIP tensor should be invariant under all symmetry operations according to Neumann's principle. These relations can be used to derive the independent tensor elements~\cite{dresselhaus_group_theory,bilbao_magnetic_group_tensor}.

\begin{table}[h]
\centering
\begin{tabular}{ cc cccccc }
\toprule
\multicolumn{2}{c}{\multirow{2}{*}{$d_{lm}(\Lambda_{lm})$}} &
\multicolumn{6}{c}{$m$} \\
& & 1 & 2 & 3 & 4 & 5 & 6  \\
\midrule
% \multicolumn{2}{c}{ $d_{lm}(\Lambda_{lm})$ } & 
% $d_{xxx}$ &  $d_{xyy}$ & $d_{xzz}$ & $d_{xyz}$ & $d_{xxz}$ & $d_{xxy}$ \\
\multirow{3}{*}{$l$} & 1 & 
$d_{xxx}$ &  $d_{xyy}$ & $d_{xzz}$ & $d_{xyz}$ & $d_{xxz}$ & $d_{xxy}$ \\
& 2 & 
$d_{yxx}$ &  $d_{yyy}$ & $d_{yzz}$ & $d_{yyz}$ & $d_{yxz}$ & $d_{yxy}$ \\
& 3 & 
$d_{zxx}$ &  $d_{zyy}$ & $d_{zzz}$ & $d_{zyz}$ & $d_{zxz}$ & $d_{zxy}$ \\
\midrule
\multicolumn{2}{c}{$\bar{\epsilon}_m$} & 
$\tilde{\epsilon}_{x}\tilde{\epsilon}_{x}$ &  $\tilde{\epsilon}_{y}\tilde{\epsilon}_{y}$ & $\tilde{\epsilon}_{z}\tilde{\epsilon}_{z}$ & $2\tilde{\epsilon}_{y}\tilde{\epsilon}_{z}$ & $2\tilde{\epsilon}_{x}\tilde{\epsilon}_{z}$ & $2\tilde{\epsilon}_{x}\tilde{\epsilon}_{y}$ \\
% \midrule

\bottomrule
\end{tabular}
\caption{Correspondence between tensor elements in Voigt and standard notation.}
\label{tab: voigt}
\end{table}

In order to derive the CIP tensor of $\alpha$-quartz, we first determine the number of independent CIP tensor elements of the $\mathrm D_3$ point group. According to group theory, this number is determined by the multiplicity of the identity representation $A_1$ contained in the direct product of $\Gamma(\bf P)$ and $\Gamma(\bar\epsilon)$ representations~\cite{dresselhaus_group_theory}. As shown in Tab.~\ref{tab: D3}, the six symmetry operations of the $D_3$ point group are divided into three classes:  $\{E\}$, $\{\mathcal C_3, \mathcal C_3^{2}\}$, and $\{\mathcal C_2^{(1)}, \mathcal C_2^{(2)}, \mathcal C_2^{(3)}\}$. The polar vectors $\tilde{\epsilon}$ and $\mathbf P$ transform according to the vector representation of $\mathrm{D}_3$, whose characters are $(3,0,-1)$. Comparing these characters with the irreducible representations listed in Tab.~\ref{tab: D3}, we obtain
\begin{eqnarray}
\label{eq: rep_ee}
\Gamma(\tilde\epsilon)\otimes\Gamma(\tilde\epsilon) &=& (E+A_2) \otimes (E+A_2) \nonumber \\
    &=& 2A_1+A_2+3E .
\end{eqnarray}
Since $\epsilon_i \epsilon_j$ is symmetric under permutation $i \leftrightarrow j$, we should only take the symmetric components of the representations in
Eq.~(\ref{eq: rep_ee}) as $\Gamma(\bar\epsilon) = [\Gamma(\tilde\epsilon)\otimes\Gamma(\tilde\epsilon)]_s$. Following the approach in Ref.~\cite{jones_group_theory}, we can obtain the characters of $\Gamma(\bar\epsilon)$ for three classes as
\begin{equation}
\begin{aligned}
\chi^{\Gamma(\bar\epsilon)}(\{E\})&=\frac{1}{2}\left[(\chi^{\Gamma(\tilde\epsilon)}(\{E\}))^2+\chi^{\Gamma(\tilde\epsilon)}(\{E\}^2)\right]=\frac{1}{2}(3^2+3)=6,\\
\chi^{\Gamma(\bar\epsilon)}(\{\mathcal C_3, \mathcal C_3^2\})&=\frac{1}{2}\left[(\chi^{\Gamma(\tilde\epsilon)}(\{\mathcal C_3, \mathcal C_3^2\}))^2+\chi^{\Gamma(\tilde\epsilon)}(\{\mathcal C_3, \mathcal C_3^2\}^2)\right]=\frac{1}{2}(0^2+0)=0,\\
\chi^{\Gamma(\bar\epsilon)}(\{\mathcal C_2^{(1)}, \mathcal C_2^{(2)}, \mathcal C_2^{(3)}\})&=\frac{1}{2}\left[(\chi^{\Gamma(\tilde\epsilon)}(\{\mathcal C_2^{(1)}, \mathcal C_2^{(2)}, \mathcal C_2^{(3)}\}))^2+\chi^{\Gamma(\tilde\epsilon)}(\{\mathcal C_2^{(1)}, \mathcal C_2^{(2)}, \mathcal C_2^{(3)}\}^2)\right]=\frac{1}{2}[(-1)^2+3]=2.\\
\end{aligned}
\end{equation}
Thus, the representation of $\Gamma(\bar\epsilon)$ is obtained as
% which can be determined by analyzing their basis functions:
\begin{eqnarray}
    \Gamma(\bar\epsilon) &=& [\Gamma(\tilde\epsilon)\otimes\Gamma(\tilde\epsilon)]_s = 2A_1 + 2E  ,
\end{eqnarray}
Finally, we can obtain
\begin{eqnarray}
    \Gamma({\bf P}) \otimes \Gamma(\bar\epsilon) &=& 
    (E+A_2) \otimes (2A_1+2E) \nonumber \\
     &=& 2A_1 + 4A_2 + 6E  .
\end{eqnarray}
Since the identity representation $A_1$ appears twice in this decomposition, the CIP tensor of $\alpha$-quartz contains two independent tensor elements under the $\mathrm{D}_3$ point group.
\begin{table}[htbp]
\centering
\begin{tabular}{c|ccc}
\hline
$D_3$ & $\{E\}$ & $\{\mathcal C_3, \mathcal C_3^{2}\}$ & $\{\mathcal C_2^{(1)}, \mathcal C_2^{(2)}, \mathcal C_2^{(3)}\}$ \\
\hline
$A_1$ & $1$ & $1$ & $1$ \\
$A_2$ & $1$ & $1$ & $-1$ \\
$E$ & $2$ & $-1$ & $0$ \\
$\Gamma(\tilde\epsilon)$, $\Gamma(\mathbf P)$ & $3$ & $0$ & $-1$ \\
$\Gamma(\bar\epsilon)$ & $6$ & $0$ & $2$ \\
\hline
\end{tabular}
\caption{The character table of $D_3$ point group with the corresponding characters for the polar vectors.}
\label{tab: D3}
\end{table}

Next, we apply the symmetry constraints of $\mathrm D_3$ group to determine the two independent CIP tensor elements.
We begin with the $\mathcal C_2^{(1)}$ operation which is a two-fold rotation along the $x$-axis:
\begin{eqnarray}
    \mathcal C_2^{(1)}: \quad (x,y,z) \to (x, -y, -z) .
\end{eqnarray}
Accordingly, the transformed vectors $\mathbf P'$ and $\tilde{\bm\epsilon}'$ should be:
\begin{eqnarray}
    \mathcal C^{(1)}_2{\mathbf P} &=& (P_x, -P_y, -P_z)=(P'_x, P'_y, P'_z ), \nonumber \\
    \mathcal C^{(1)}_2\tilde{\bm\epsilon} &=& (\tilde{\epsilon}_x, -\tilde{\epsilon}_y, -\tilde{\epsilon}_z)=(\tilde{\epsilon}'_x, \tilde{\epsilon}'_y, \tilde{\epsilon}'_z ).
    \label{eq: P after C2 rotation}
\end{eqnarray}
Therefore, combining these two expressions we have:
\begin{eqnarray}
    P'_x = P_x &=&  d_{11} \bar\epsilon_1 + d_{12} \bar\epsilon_2 + d_{13} \bar\epsilon_3 + d_{14} \bar\epsilon_4 + d_{15} \bar\epsilon_5 + d_{16} \bar\epsilon_6
    \nonumber \\
     &\stackrel{!}{=}& d_{11} \bar\epsilon_1 + d_{12} \bar\epsilon_2 + d_{13} \bar\epsilon_3 + d_{14} \bar\epsilon_4 - d_{15} \bar\epsilon_5 - d_{16} \bar\epsilon_6 = d_{1m} \bar{\epsilon}'_m, \nonumber \\
    P'_y = -P_y &=& - d_{21} \bar\epsilon_1 - d_{22} \bar\epsilon_2 - d_{23} \bar\epsilon_3 - d_{24} \bar\epsilon_4 - d_{25} \bar\epsilon_5 - d_{26} \bar\epsilon_6
    \nonumber \\
     &\stackrel{!}{=}& d_{21} \bar\epsilon_1 + d_{22} \bar\epsilon_2 + d_{23} \bar\epsilon_3 + d_{24} \bar\epsilon_4 - d_{25} \bar\epsilon_5 - d_{26} \bar\epsilon_6 = d_{2m}\bar{\epsilon}_m', \nonumber \\
    P'_z = -P_z &=& -d_{31} \bar\epsilon_1 - d_{32} \bar\epsilon_2 - d_{33} \bar\epsilon_3 - d_{34} \bar\epsilon_4 - d_{35} \bar\epsilon_5 - d_{36} \bar\epsilon_6
    \nonumber \\
     &\stackrel{!}{=}& d_{31} \bar\epsilon_1 + d_{32} \bar\epsilon_2 + d_{33} \bar\epsilon_3 + d_{34} \bar\epsilon_4 - d_{35} \bar\epsilon_5 - d_{36} \bar\epsilon_6 = d_{3m}\bar{\epsilon}'_m ,
\end{eqnarray}
where $\stackrel{!}{=}$ means `required to be equal to'.
From these relations, many tensor elements $d_{lm}$ must vanish:
\begin{eqnarray}
    d_{15} &=& 0, \ d_{16} = 0, \nonumber \\
    d_{21} &=& 0, \ d_{22} = 0, \ d_{23} = 0, \ d_{24} = 0, \nonumber \\
    d_{31} &=& 0, \ d_{32} = 0, \ d_{33} = 0, \ d_{34} = 0. \nonumber
\end{eqnarray}
Therefore, after applying $C_2^{(1)}$, the CIP tensor is reduced to eight independent elements:
\begin{eqnarray}
    d = \begin{pmatrix}
        d_{11} & d_{12} & d_{13} & d_{14} & 0 & 0 \\
        0 & 0 & 0 & 0 & d_{25} & d_{26} \\
        0 & 0 & 0 & 0 & d_{35} & d_{36} 
    \end{pmatrix} \ .
\end{eqnarray}

Next, we consider $\mathcal C_3$ operation, which rotates the $x$ and $y$ axes by 120 degrees:
% \begin{eqnarray}
%     C_3: \quad (x,y,z) \to (-\frac{1}{2}x+\frac{\sqrt{3}}{2}y, -\frac{\sqrt{3}}{2}x-\frac{1}{2}y, z) \nonumber \\
% \end{eqnarray}
\begin{eqnarray}
    \mathcal C_3{\mathbf P} &=& \left(-\frac{1}{2}P_x+\frac{\sqrt{3}}{2}P_y, -\frac{\sqrt{3}}{2}P_x-\frac{1}{2}P_y, P_z\right) = (P'_x, P'_y, P'_z), \nonumber \\
    \mathcal C_3 {\tilde{\bm\epsilon}} &=& \left(-\frac{1}{2}\tilde\epsilon_x+\frac{\sqrt{3}}{2}\tilde\epsilon_y, -\frac{\sqrt{3}}{2}\tilde\epsilon_x-\frac{1}{2}\tilde\epsilon_y, \tilde\epsilon_z\right) = (\tilde\epsilon'_x, \tilde\epsilon'_y, \tilde\epsilon'_z). \nonumber
\end{eqnarray}
Therefore, on one hand we have:
\begin{eqnarray}
    P'_x &=& d_{1m}\bar{\epsilon}'_m = d_{11} \tilde\epsilon'_x\tilde\epsilon'_x + d_{12} \tilde\epsilon'_y\tilde\epsilon'_y + d_{13} \tilde\epsilon'_z\tilde\epsilon'_z + d_{14} \cdot 2\tilde\epsilon'_y\tilde\epsilon'_z
    \nonumber \\
    &=& d_{11} \left(-\frac{1}{2}\tilde\epsilon_x+\frac{\sqrt{3}}{2}\tilde\epsilon_y\right)^2 + d_{12} \left(-\frac{\sqrt{3}}{2}\tilde\epsilon_x-\frac{1}{2} \tilde\epsilon_y\right)^2 + d_{13} \tilde\epsilon_z \tilde\epsilon_z 
    %\nonumber \\
    + d_{14}\cdot2\left(-\frac{\sqrt{3}}{2}\tilde\epsilon_x-\frac{1}{2} \tilde\epsilon_y\right)\tilde\epsilon_z \nonumber \\
    &=& \left(\frac{1}{4}d_{11} + \frac{3}{4}d_{12}\right) \bar\epsilon_1 + \left(\frac{3}{4} d_{11} + \frac{1}{4} d_{12}\right) \bar\epsilon_2 + d_{13} \bar\epsilon_3 - \frac{1}{2} d_{14} \bar\epsilon_4 
    %\nonumber \\
    -\frac{\sqrt{3}}{2}d_{14}\bar\epsilon_5 + \left(-\frac{\sqrt{3}}{4}d_{11} + \frac{\sqrt{3}}{4} d_{12}\right)\bar\epsilon_6
    \nonumber \\
    P'_y &=& d_{2m}\bar{\epsilon}'_m = d_{25} \cdot 2 \left(-\frac{1}{2}\tilde\epsilon_x+\frac{\sqrt{3}}{2}\tilde\epsilon_y\right)\tilde\epsilon_z 
    %\nonumber \\
    + d_{26} \cdot 2 \left(-\frac{1}{2}\tilde\epsilon_x+\frac{\sqrt{3}}{2}\tilde\epsilon_y\right) \left(-\frac{\sqrt{3}}{2}\tilde\epsilon_x-\frac{1}{2} \tilde\epsilon_y\right) \nonumber \\
    &=& \frac{\sqrt{3}}{2}d_{26} \bar\epsilon_1 - \frac{\sqrt{3}}{2} d_{26} \bar\epsilon_2 + \frac{\sqrt{3}}{2}d_{25} \bar\epsilon_4 - \frac{1}{2} d_{25}\bar\epsilon_5 - \frac{1}{2}d_{26} \bar \epsilon_6
    \nonumber \\
    P'_z &=& d_{3m}\bar{\epsilon}'_m =  d_{35} \cdot 2 \left(-\frac{1}{2}\tilde\epsilon_x+\frac{\sqrt{3}}{2}\tilde\epsilon_y\right)\tilde\epsilon_z
    %\nonumber \\
    + d_{36} \cdot 2\left(-\frac{1}{2}\tilde\epsilon_x+\frac{\sqrt{3}}{2}\tilde\epsilon_y\right) \left(-\frac{\sqrt{3}}{2}\tilde\epsilon_x-\frac{1}{2} \tilde\epsilon_y\right) \nonumber \\
    &=& \frac{\sqrt{3}}{2}d_{36} \bar\epsilon_1 - \frac{\sqrt{3}}{2} d_{36} \bar\epsilon_2 + \frac{\sqrt{3}}{2}d_{35} \bar\epsilon_4 - \frac{1}{2} d_{35}\bar\epsilon_5 - \frac{1}{2}d_{36} \bar \epsilon_6
    \label{eq: C3 LHS}
\end{eqnarray}
In the above derivation, note the factor of 2 in front of some components in the Voigt notation, like $\bar\epsilon_4 = 2\epsilon_y\epsilon_z$.
And on the other hand, we also have:
\begin{eqnarray}
    P'_x &=& -\frac{1}{2}P_x+\frac{\sqrt{3}}{2}P_y = 
    -\frac{1}{2} d_{11}\bar\epsilon_1 -\frac{1}{2} d_{12}\bar\epsilon_2
     -\frac{1}{2} d_{13}\bar\epsilon_3  
     %\nonumber \\
     -\frac{1}{2} d_{14}\bar\epsilon_4  
     + \frac{\sqrt{3}}{2} d_{25}\bar\epsilon_5 
    + \frac{\sqrt{3}}{2} d_{26}\bar\epsilon_6 
    \nonumber \\
    P'_y &=&
    -\frac{\sqrt{3}}{2}P_x-\frac{1}{2} P_y = 
    -\frac{\sqrt{3}}{2} d_{11}\bar\epsilon_1 -\frac{\sqrt{3}}{2} d_{12}\bar\epsilon_2
     -\frac{\sqrt{3}}{2} d_{13}\bar\epsilon_3  
     %\nonumber \\
     -\frac{\sqrt{3}}{2} d_{14}\bar\epsilon_4 
     - \frac{1}{2} d_{25}\bar\epsilon_5 
    - \frac{1}{2} d_{26}\bar\epsilon_6
    \nonumber \\
    P'_z &=& P_z = d_{35}\bar{\epsilon}_5 +  d_{36}\bar{\epsilon}_6
    \label{eq: C3 RHS}
\end{eqnarray}
Comparing Eqs.~(\ref{eq: C3 LHS}) and Eqs.~(\ref{eq: C3 RHS}), we can further eliminate some tensor elements:
\begin{eqnarray}
    d_{11} &=& -d_{12}, \  d_{13} = 0, \ d_{14} = -d_{25}, \ d_{11} = -d_{26} 
    %\nonumber \\
    \ , \ d_{35} = 0, \ d_{36} = 0 . \nonumber
\end{eqnarray}
Since now we are left with only two independent CIP tensor elements consistent with the number of independent elements derived above, the CIP tensor of point group D$_3$ is:
\begin{eqnarray}
    d = \begin{pmatrix}
        d_{11} & -d_{11} & 0 & d_{14} & 0 & 0 \\
        0 & 0 & 0 & 0 & -d_{14} & -d_{11} \\
        0 & 0 & 0 & 0 & 0 & 0 
    \end{pmatrix} \ .
    \label{eq: D3 cip tensor app}
\end{eqnarray}
Note that the CIP tensor, like any response tensor, depends on the choice of Cartesian coordinate system. Here we choose the axes such that one of the $\mathcal C^{(1)}_2$ symmetry axes lies along the $x$ direction. If instead a $\mathcal C^{(1)}_2$ axis is aligned with $y$, the corresponding CIP tensor is obtained by interchanging $x \leftrightarrow y$ in Eq.~(\ref{eq: D3 cip tensor app}).

All CIP tensors for the ten nonpolar point groups that are compatible with the CIP effect, together with representative materials, are listed in Tab.~\ref{tab:all CIP tensors}.
\begin{table*}[]
    \centering
    \begin{tabular}{c c c }
    \toprule
PGs & CIP Tensor & material  \\
\midrule
$\mathrm{D_2}$ &
$
\begin{pmatrix}
        0 & 0 & 0 & d_{14} & 0 & 0 \\
        0 & 0 & 0 & 0 & d_{25} & 0 \\
        0 & 0 & 0 & 0 & 0 & d_{36}
\end{pmatrix}
$
& AlPS$_4$ (mp-27462) \\
\\
$\mathrm{D}_{2d}$ &
$
\begin{pmatrix}
        0 & 0 & 0 & d_{14} & 0 & 0 \\
        0 & 0 & 0 & 0 & d_{14} & 0 \\
        0 & 0 & 0 & 0 & 0 & d_{36}
\end{pmatrix}
$
& AgClO$_4$ (mp-22993) \\
\\
$\mathrm{D}_{3}$ &
$
\begin{pmatrix}
        d_{11} & -d_{11} & 0 & d_{14} & 0 & 0 \\
        0 & 0 & 0 & 0 & -d_{14} & -d_{11} \\
        0 & 0 & 0 & 0 & 0 & 0
\end{pmatrix}
$
& $\alpha$-quartz (mp-6930) \\
\\
$\mathrm{C}_{3h}$ &
$
\begin{pmatrix}
        d_{11} & -d_{11} & 0 & 0 & 0 & -d_{22} \\
        -d_{22} & d_{22} & 0 & 0 & 0 & -d_{11} \\
        0 & 0 & 0 & 0 & 0 & 0
\end{pmatrix}
$
& BaZnBO$_3$F (mp-1077866) \\
\\
$\mathrm{D}_{3h}$ &
$
\begin{pmatrix}
        0 & 0 & 0 & 0 & 0 & -d_{22} \\
        -d_{22} & d_{22} & 0 & 0 & 0 & 0 \\
        0 & 0 & 0 & 0 & 0 & 0
\end{pmatrix}
$
& MoS$_2$ (mp-1023924) \\
\\
$\mathrm{S}_{4}$ &
$
\begin{pmatrix}
        0 & 0 & 0 & d_{14} & d_{15} & 0 \\
        0 & 0 & 0 & -d_{15} & d_{14} & 0 \\
        d_{31} & -d_{31} & 0 & 0 & 0 & d_{36}
\end{pmatrix}
$
& BPO$_4$ (mp-3589) \\
\\
$\mathrm{D}_{4}$ &
$
\begin{pmatrix}
        0 & 0 & 0 & d_{14} & 0 & 0 \\
        0 & 0 & 0 & 0 & -d_{14} & 0 \\
        0 & 0 & 0 & 0 & 0 & 0
\end{pmatrix}
$
& Cristobalite (mp-6945) \\
\\
$\mathrm{D}_{6}$ &
$
\begin{pmatrix}
        0 & 0 & 0 & d_{14} & 0 & 0 \\
        0 & 0 & 0 & 0 & -d_{14} & 0 \\
        0 & 0 & 0 & 0 & 0 & 0
\end{pmatrix}
$
& $\beta$-quartz (mp-6922) \\
\\
$\mathrm{T}$ &
$
\begin{pmatrix}
        0 & 0 & 0 & d_{14} & 0 & 0 \\
        0 & 0 & 0 & 0 & d_{14} & 0 \\
        0 & 0 & 0 & 0 & 0 & d_{14}
\end{pmatrix}
$
%& SbIrS (mp-8630) \\
& K$_3$SbS$_3$ (mp-1194266) \\
\\
$\mathrm{T}_{d}$ &
$
\begin{pmatrix}
        0 & 0 & 0 & d_{14} & 0 & 0 \\
        0 & 0 & 0 & 0 & d_{14} & 0 \\
        0 & 0 & 0 & 0 & 0 & d_{14}
\end{pmatrix}
$
& BP (mp-1479) \\
\bottomrule
    \end{tabular}
    \caption{Corresponding CIP tensors and representative materials (with Materials Project~\cite{materials_project} IDs) for ten nonpolar point groups compatible with the CIP effect.}
    \label{tab:all CIP tensors}
\end{table*}

\section{Computational details}

Group-theoretical analysis in this work is carried out with the aid of tables and programs on the Bilbao Crystallographic Server~\cite{bilbao_paper,bilbao_magnetic_group_tensor,bilbao_spin_group_tensor}. The spin-group symmetry operations of Mn$_3$Sn are obtained with the help of  FINDSPINGROUP program~\cite{qihang_prx_spin_group_2024}. 

\begin{center}
    \begin{figure}[h]
        \centering
        \includegraphics[width=0.8\linewidth]{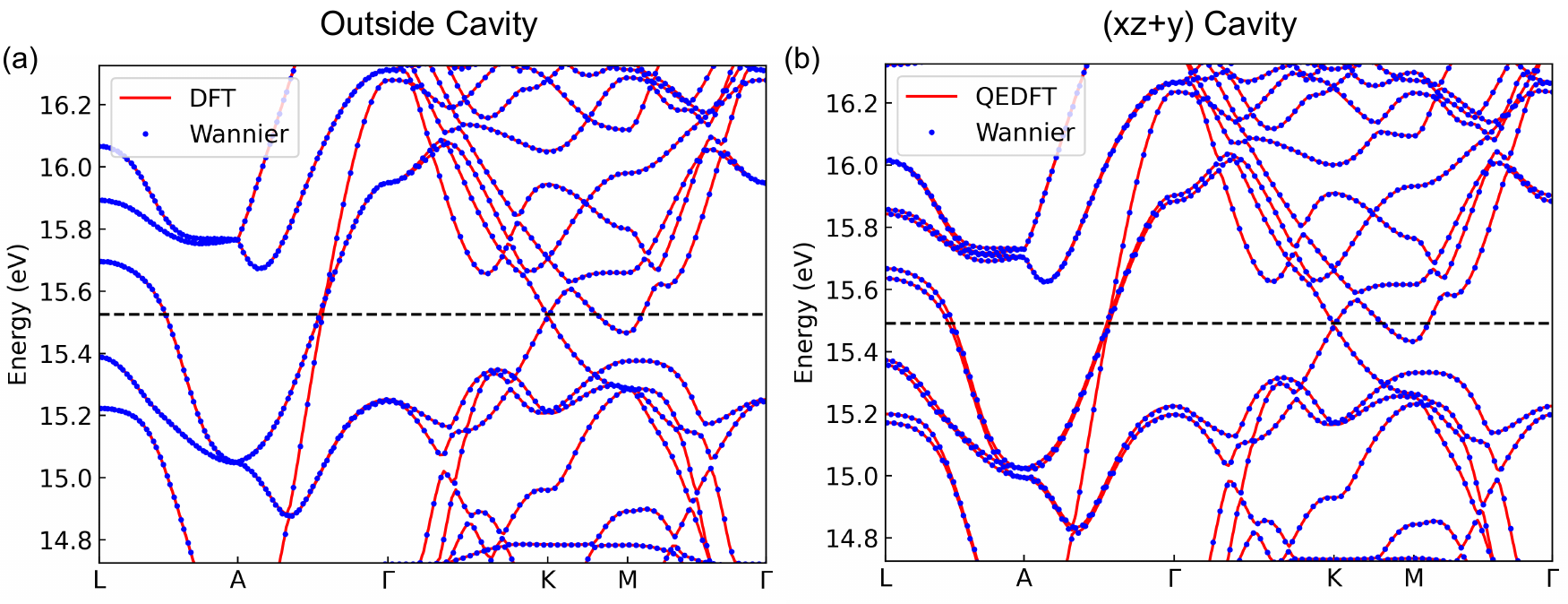}
        \caption{ (a) Band structures of AFM-1 Mn$_3$Sn outside the cavity, calculated from DFT (red lines) and Wannier interpolation (blue dots). (b) Band structures of AFM-1 Mn$_3$Sn coupled to cavity modes in the ($xz+y$) plane from QEDFT (red lines) and Wannier interpolation (blue dots). The fractional coordinates of high symmetry points are: L$(0.5, 0.0, 0.5)$, A$(0.0, 0.0, 0.5)$, $\Gamma(0.0,0.0,0.0)$, K$(1/3, 1/3, 0.0)$, and M$(0.5, 0.0, 0.0)$.
        }
        \label{fig:Mn3Sn wannier}
    \end{figure}
\end{center}

All density functional theory (DFT) and QEDFT calculations are performed using our in-house implementation of the pxLDA functional~\cite{I-te_pxLDA,benshu_qedft} within the {\tt Quantum ESPRESSO} package~\cite{qe_2020}. SiO$_2$ ($\alpha$-quartz) and Mn$_3$Sn are calculated using scalar relativistic and fully relativistic optimized norm-conserving Vanderbilt pseudopotentials~\cite{ONCV_pseudopotential,pseudodojo}, respectively. The numbers of valence electrons in the unit cell are $N_e = 48$ and 118 for $\alpha$-quartz and Mn$_3$Sn, respectively.
The Perdew-Burke-Ernzerhof (PBE) electron-electron exchange correlation functional~\cite{pbe} is employed for both materials. 
Note that these bare photon mode parameters are transformed into the dressed photon mode $\tilde{\bm \epsilon}$ and $\tilde{\lambda}/\tilde{\omega}$ by Bogoliubov transformation~\cite{I-te_pxLDA} within the code. 
For $\alpha$-quartz, self-consistent calculations both with and without cavity photon modes are performed using
a kinetic energy cutoff of 80 Ry for the wavefunctions and a $10\times 10\times 10$ $\bf k$-grid for Brillouin zone (BZ) sampling. In the QEDFT calculation, ${\bm \epsilon} = (1,0,0), (0,1,0), (0,0,1), 1/\sqrt{2}(1,1,0), 1/\sqrt{2}(0,1,1)$, and $1/\sqrt{2}(1,0,1)$ are used to represent the $x$-, $y$-, $z$-, $xy$-, $yz$- and $xz$-polarized cavity modes. Ten evenly spaced values of $\lambda/\omega$ in the interval $[0.01, 0.10]$ are adopted to compute the induced polarization. The angle-dependent polarizations in the $xy$ plane are calculated with two cavity photon modes ${\bm \epsilon}=(\cos\theta, \sin\theta,0)+(0,0,1)$, where $\theta$ indicates the angle with the $x$ axis. The polarization is evaluated using the Berry phase approach~\cite{king_vanderbilt_modern_polarization,vanderbilt_berry_phase_book}. 
Computational details of the other nine CIP materials are given in Sec.~\ref{app: ten cip materials} below.
For Mn$_3$Sn, a kinetic energy cutoff of 96 Ry for the wavefunctions and a $9\times 9\times 10$ $\bf k$-grid are employed for self-consistent calculations both with and without cavity photon modes.  The spin-orbit coupling (SOC) effect is included in all Mn$_3$Sn calculations in the main text, and the results without SOC are discussed in Sec.~\ref{app: cim SPG tensor} below. For the QEDFT calculations, seven evenly spaced values of ${{\lambda}}/{{\omega}}$ in the interval $[0.03, 0.20]$ are chosen to compute the total magnetization. 
The initial magnetization directions on the Mn atoms are set according to the configuration shown in Fig. 3(a) of the main text, while both their magnitudes and directions are relaxed without constraints during the self-consistent procedure. The anomalous Hall conductivity component $\sigma_{ij}$ is calculated by:
\begin{eqnarray}
    \sigma_{ij} = -\frac{e^2}{(2\pi)^3\hbar}\sum_n\int_{BZ} f_{n{\bf k}}\Omega_{n,ij}({\bf k}) \ d^3{\bf k},
\end{eqnarray}
as implemented in the {\tt Wannier90} package~\cite{Wang_ahc_wannier, wannier90_jpcm}. Here $n$ is the band index, $f_{n{\bf k}}$ and $\Omega_{n,ij}({\bf k})$ are the occupation number and the Berry curvature of electronic eigenstate $\ket{n{\bf k}}$.
The Brillouin zone integration is calculated on a dense $150\times150\times150$ $\bf k$-grid via Wannier interpolation.  For Wannierization, we use a $\bf k$-grid of $10 \times 10\times 10$ to calculate non-self-consistent wavefunctions and choose Mn-$d$ orbitals and Sn-$p$ orbitals as projectors. The band structures obtained from Wannier interpolation are shown in Fig.~\ref{fig:Mn3Sn wannier}, which demonstrate the high quality of the Wannierization.

\section{QEDFT calculations of ten CIP materials}
\label{app: ten cip materials}

\begin{center}
    \begin{figure*}[ht!]
        \centering
        \includegraphics[width=1\linewidth]{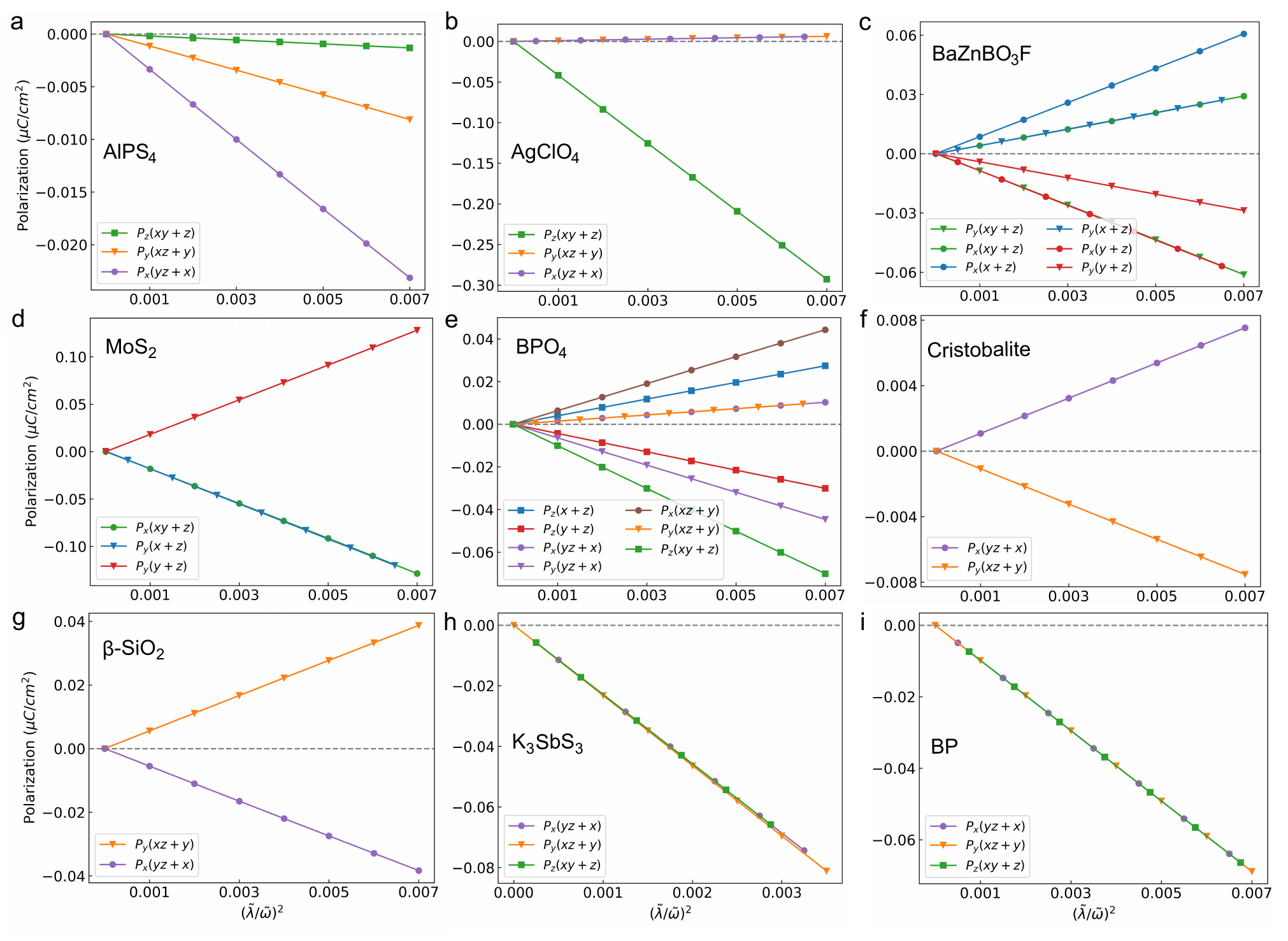}
        \caption{ Cavity-induced polarization produced by various cavity mode configurations computed with QEDFT for the materials listed in Tab.~\ref{tab:all CIP tensors}. All the results are fully compatible with the derived CIP tensor. The results of $\alpha$-quartz are discussed in the main text.
        }
        \label{fig:ten_cip}
    \end{figure*}
\end{center}

Here we present QEDFT-calculated cavity-induced polarizations for representative materials in the remaining nine CIP-compatible nonpolar point groups in Fig.~\ref{fig:ten_cip}. All the results are fully consistent with the corresponding CIP tensors summarized in Tab.~\ref{tab:all CIP tensors}. For these nine materials, we employ optimized norm-conserving Vanderbilt pseudopotentials~\cite{ONCV_pseudopotential,pseudodojo} together with the PBE~\cite{pbe} functional. Prior to the polarization calculations, all the atoms are fully relaxed until the residual forces are below $10^{-4}$ Ry/Bohr with lattice constants fixed at their pristine values. Additional computational parameters are provided in Tab.~\ref{tab:details ten cip}. 

\begin{table*}[]
    \centering
    \begin{tabular}{c c c c }
    \toprule
material & $\bf k$-grid & energy cutoff (Ry) & $N_e$ \\
\midrule
AlPS$_4$ & $8 \times 8\times 5$ 
& 52 & 64\\
\\
AgClO$_4$ & $9 \times 9\times 7$
& 84 & 100 \\
\\
BaZnBO$_3$F & $10 \times 10\times 10$
& 84 & 58 \\
\\
MoS$_2$ &  $16 \times 16\times 1$
& 80 & 26\\
\\
BPO$_4$ & $10 \times 10\times 7$
& 84 & 64 \\
\\
Cristobalite & $9 \times 9\times 7$
& 84 & 64\\
\\
$\beta$-quartz & $10 \times 10\times 8$
&  84 & 48\\
\\
K$_3$SbS$_3$ & $5 \times 5\times 5$
& 80 & 240\\
\\
BP & $10 \times 10\times 10$
& 76 & 32\\
\bottomrule
    \end{tabular}
    \caption{Computational parameters in the self-consistent calculation for representative materials in the nine remaining CIP-compatible point groups.}
    \label{tab:details ten cip}
\end{table*}

\section{Derivation of the independent CIM tensor elements with magnetic group symmetry}
\label{app: cim MPG tensor}
The CIM effect can be written in Voigt notation as:
\begin{eqnarray}
    \begin{pmatrix}
        M_x \\
        M_y \\
        M_z 
    \end{pmatrix} =   \begin{pmatrix}
        \Lambda_{11} & \Lambda_{12} & \Lambda_{13} & \Lambda_{14} & \Lambda_{15} & \Lambda_{16} \\
        \Lambda_{21} & \Lambda_{22} & \Lambda_{23} & \Lambda_{24} & \Lambda_{25} & \Lambda_{26} \\
        \Lambda_{31} & \Lambda_{32} & \Lambda_{33} & \Lambda_{34} & \Lambda_{35} & \Lambda_{36} 
    \end{pmatrix} 
    \begin{pmatrix}
        \bar\epsilon_{1} \\
        \bar\epsilon_{2} \\
        \bar\epsilon_{3} \\
        \bar\epsilon_{4} \\
        \bar\epsilon_{5} \\
        \bar\epsilon_{6}
    \end{pmatrix}.
    \label{eq: CIM general}
\end{eqnarray}
Here we derive the CIM tensor of coplanar AFM-1 Mn$_3$Sn, which corresponds to the MPG $mm'm'$. 
Note that under a MPG symmetry operation, spin (i.e., local magnetic moment) and atom transform simultaneously. However, it is not the case for spin group symmetries, which will be discussed in Sec.~\ref{app: cim SPG tensor}.  We first consider $\mathcal{C}^{(1)}_2$ along $x$ axis in MPG $mm'm'$. 
Under $\mathcal{C}^{(1)}_2$ operation $\tilde{\bm \epsilon}$ and $\bm M$ change as:
\begin{eqnarray}
    \mathcal{C}^{(1)}_2 {\bm M} &=& (M_x, -M_y, -M_z) = (M'_x, M'_y, M'_z) \nonumber \\
     \mathcal{C}^{(1)}_2 {\tilde{\bm \epsilon}} &=& (\tilde{\epsilon}_x, -\tilde{\epsilon}_y, -\tilde{\epsilon}_z) = (\tilde{\epsilon}'_x, \tilde{\epsilon}'_y, \tilde{\epsilon}'_z).
\end{eqnarray}
Similar to the procedure given in Sec.~\ref{app: cip tensor}, we get:
\begin{eqnarray}
            \Lambda = 
    \begin{pmatrix}
        \Lambda_{11} & \Lambda_{12} & \Lambda_{13} & \Lambda_{14} & 0 & 0 \\
        0 & 0 & 0 & 0 & \Lambda_{25} & \Lambda_{26} \\
        0 & 0 & 0 & 0 & \Lambda_{35} & \Lambda_{36} 
    \end{pmatrix}.
    \label{eq: CIM MPG C2}
\end{eqnarray}
Next we take operation $\mathcal{M}_y\mathcal{T}$ in  $mm'm'$, which transforms $\tilde{\bm \epsilon}$ and $\bm M$ as:
\begin{eqnarray}
        \mathcal{M}_y\mathcal{T} {\bm M} &=& (M_x, -M_y, M_z) = (M'_x, M'_y, M'_z)\nonumber \\
     \mathcal{M}_y\mathcal{T} {\tilde{\bm \epsilon}} &=& (\tilde{\epsilon}_x, -\tilde{\epsilon}_y, \tilde{\epsilon}_z)=(\tilde{\epsilon}'_x,\tilde{\epsilon}'_y, \tilde{\epsilon}'_z).
\end{eqnarray}
Comparing with Eq.~(\ref{eq: CIM MPG C2}), we have:
\begin{eqnarray}
    M'_x = M_x &=& \Lambda_{11}\bar{\epsilon}_1 + \Lambda_{12}\bar{\epsilon}_2 + \Lambda_{13}\bar{\epsilon}_3 + \Lambda_{14}\bar{\epsilon}_4 \nonumber \\
    &\stackrel{!}{=}&  \Lambda_{11}\bar{\epsilon}_1 + \Lambda_{12}\bar{\epsilon}_2 + \Lambda_{13}\bar{\epsilon}_3 - \Lambda_{14}\bar{\epsilon}_4
    = \Lambda_{1m} \bar{\epsilon}_m'\\
    M'_y =-M_y &=& -\Lambda_{25}\bar{\epsilon}_5 - \Lambda_{26}\bar{\epsilon}_6 \nonumber \\
    &\stackrel{!}{=}&  \Lambda_{25}\bar{\epsilon}_5 - \Lambda_{26}\bar{\epsilon}_6 =\Lambda_{2m} \bar{\epsilon}_m' \\
    M'_z = M_z &=& \Lambda_{35}\bar{\epsilon}_5 + \Lambda_{36}\bar{\epsilon}_6 \nonumber \\
    &\stackrel{!}{=}&  \Lambda_{35}\bar{\epsilon}_5 - \Lambda_{36}\bar{\epsilon}_6 =\Lambda_{3m} \bar{\epsilon}_m' .
\end{eqnarray}
As a result, the following CIM elements vanish:
\begin{eqnarray}
    \Lambda_{14} = 0, \Lambda_{25}=0, \Lambda_{36} = 0.
\end{eqnarray}
Applying other operations does not further reduce the elements. Therefore, the CIM tensor of MPG $mm'm'$ is:
\begin{eqnarray}
            \Lambda = 
    \begin{pmatrix}
        \Lambda_{11} & \Lambda_{12} & \Lambda_{13} & 0 & 0 & 0 \\
        0 & 0 & 0 & 0 & 0 & \Lambda_{26} \\
        0 & 0 & 0 & 0 & \Lambda_{35} & 0 
    \end{pmatrix}.
\end{eqnarray}
Following this procedure, one can derive the CIM tensors for all MPGs listed in Tab.~\ref{tab: cim full pg}. Since there are 66 CIM-compatible MPGs, we do not present the CIM tensors for all groups here for brevity. 

\section{Connection to experimental setup}
\label{app: exp cavity setup}

{
In experiments, the light-matter coupling strength $\lambda$ and the mode frequency $\omega$ highly depend on the explicit type of the cavity. In this section, we briefly discuss the relation between the dressed coupling ratio $\tilde{\lambda}/\tilde{\omega}$, Fabry-P{\'e}rot cavity setup, and the sample size in experiments. 
Because for a Fabry-P{\'e}rot cavity, both the coupling strength and cavity mode frequency are directly related to the cavity mirror separation distance, whereas for other types of cavity, these relations are generally unclear.
For the cases with one effective cavity mode, the dressed coupling ratio is related to the bare collective light-matter coupling strength $\lambda$ and frequency $\omega$ through
\begin{eqnarray}
    \frac{\tilde\lambda^2}{\tilde{\omega}^2} = \frac{\lambda^2}{\omega^2+N_e\lambda^2}.
\end{eqnarray}
Within the effective photon-free QEDFT framework, the bare collective coupling strength can be written as $\lambda=\sqrt{N_{\rm cell}} \lambda^{\rm uc}$, where $\lambda^{\rm uc}$ is the coupling strength per unit cell~\cite{benshu_qedft}. This scaling is justified by the long-wavelength approximation, under which the cavity photon field is spatially uniform on the
scale of the crystal and therefore couples identical to the current in each unit cell. Consequently, each unit cell contributes coherently to
the light-matter interaction. The unit cell coupling strength $\lambda^{\rm uc}$ is determined by the cavity mode volume $V_{\rm mode}$ via $\lambda^{\rm uc}=\sqrt{4\pi/V_{\rm mode}}$, where $V_{\rm mode}$ can be estimated as $V_{\rm mode}=L^3_c \mathcal{F}$. Here, $L_c$ is the cavity mirror separation and $\mathcal{F}$ is the cavity finesse, which is related to the reflection coefficient $r$ through
$\mathcal{F}=-2\pi/\ln(|r|^4)$~\cite{Mark_effective_single_mode_2025}. Combining these expressions, the dressed coupling ratio can be written as
\begin{eqnarray}
\label{eq: dressed to mode volume}
        \frac{\tilde\lambda^2}{\tilde{\omega}^2} = \frac{4\pi N_{\rm cell}}{\omega^2V_{\rm mode}+4\pi N_{\rm cell}N_e}.
\end{eqnarray}
As an example, for Mn$_3$Sn, taking $r=0.8$ and $L_c = 0.5$ $\mu$m gives a mode volume $V_{\rm mode}\sim 5.9\times10^{12}$ Bohr$^3$. Using the fundamental cavity mode frequency $\omega= \pi c/L_c=0.0456$ Ha and $N_e =118$, we can solve Eq.~(\ref{eq: dressed to mode volume}) to get, for instance, the corresponding $N_{\rm cell}\sim 340^3$ for the dressed coupling ratio $\tilde{\lambda}^2/\tilde{\omega}^2=0.007$. Since the lattice vector of Mn$_3$Sn is approximately 5 \AA, this corresponds to a sample size of roughly 0.17 $\times$ 0.17 $\times$ 0.17 $\mu$m$^3$.
}

\section{Derivation of the independent CIM tensor elements with spin group symmetry}
\label{app: cim SPG tensor}

Here we derive the CIM tensor of coplanar AFM-1 Mn$_3$Sn without SOC, which is instead described by the spin point group. 
In general, a spin point group ($G_{\rm{SP}}$) is given by the direct product of a spin-only group ($G_{\rm{SO}}$) and a nontrivial spin group ($G_{\rm{NS}}$), i.e., $G_{\rm SP} = G_{\rm SO} \otimes G_{\rm NS}$~\cite{litvin_spin_point_group,libor_prx_altermagnet_2022,qihang_prx_spin_group_2022}. For any coplanar spin arrangement, $G_{\rm SO} = \{E, {\mathcal T}2_{001}\}$~\cite{qihang_prx_spin_group_2022}, where $2_{001}$ is the twofold rotation in spin space about the axis perpendicular to the Kagome plane. For AFM-1 Mn$_3$Sn, the nontrivial spin group is $G_{\rm NS}\ {=}\ ^{3_z}6/^1m^{2_x}m^{2_{xy}}m$ ($\# 498$ in Litvin's table~\cite{litvin_spin_point_group} and category-IX in Ref.~\cite{qihang_prx_spin_group_2022}), where we use Litvin's notation for spin groups.
Therefore, the spin point group of AFM-1 Mn$_3$Sn is $G_{SO}\otimes G_{NS} = \{E,\mathcal{T}2_{001}\} \otimes \ ^{3_z}6/^1m^{2_x}m^{2_{xy}}m$. To derive the corresponding CIM tensor elements, we first consider the symmetry operation $\{ \mathcal{T}2_{001}|| E\}$ in the spin-only group.
Here we adopt the notation $\{ r_s||R_s\}$
for spin point group symmetry operations, where the operation $r_s$ on the left acts in spin space, while $R_s$ on the right acts in real space~\cite{litvin_spin_point_group,libor_prx_altermagnet_2022}.
 
Under this operation, $\bm M$ and $\tilde{\bm \epsilon}$ change as:
\begin{eqnarray}
    \{ \mathcal{T}2_{001}|| E\} {\bm M} &=& (M_x, M_y, -M_z) = (M'_x, M'_y, M'_z), \nonumber \\
     \{ \mathcal{T}2_{001}|| E\}{\tilde{\bm \epsilon}} &=& (\tilde\epsilon_x, \tilde\epsilon_y, \tilde\epsilon_z) = (\tilde\epsilon'_x, \tilde\epsilon'_y, \tilde\epsilon'_z).
\end{eqnarray}
Note that $\bm M$ only changes under spin-space operations, because $\bm M$ we considered here is entirely contributed by spin.
According to Neumann's principle, tensor $\Lambda$ should be invariant under all symmetry operations in this group. This requires $M'_z=-M_z = \Lambda_{3m}\bar{\epsilon}_m = M_z$, which can only be fulfilled if all $\Lambda_{3m} = 0$, hence:
\begin{eqnarray}
    \Lambda = 
    \begin{pmatrix}
        \Lambda_{11} & \Lambda_{12} & \Lambda_{13} & \Lambda_{14} & \Lambda_{15} & \Lambda_{16} \\
        \Lambda_{21} & \Lambda_{22} & \Lambda_{23} & \Lambda_{24} & \Lambda_{25} & \Lambda_{26} \\
        0 & 0 & 0 & 0 & 0 & 0 
    \end{pmatrix} .
\end{eqnarray}
Next, we apply the symmetry operation $\{2_{100}||2_{100}\}$ in $G_{NS}$ to the CIM tensor. Under this operation:
\begin{eqnarray}
    \{E||2_{100}\}{\tilde{\bm \epsilon}} &=& (\tilde\epsilon_x, -\tilde\epsilon_y, -\tilde\epsilon_z) = (\tilde\epsilon'_x, \tilde\epsilon'_y, \tilde\epsilon'_z),\nonumber \\
    \{2_{100}||E\}{\bm M} &=& (M_x, -M_y, -M_z) = (M'_x, M'_y, M'_z).
\end{eqnarray}
This leads to:
\begin{eqnarray}
    M'_x = M_x &=&
    \Lambda_{11} \bar{\epsilon}_1 + \Lambda_{12} \bar{\epsilon}_2 + \Lambda_{13} \bar{\epsilon}_3 + \Lambda_{14} \bar{\epsilon}_4 + \Lambda_{15} \bar{\epsilon}_5 +\Lambda_{16} \bar{\epsilon}_6
    \nonumber \\
    &\stackrel{!}{=}& \Lambda_{11} \bar{\epsilon}_1 + \Lambda_{12} \bar{\epsilon}_2 + \Lambda_{13} \bar{\epsilon}_3 + \Lambda_{14} \bar{\epsilon}_4 - \Lambda_{15} \bar{\epsilon}_5 -\Lambda_{16} \bar{\epsilon}_6 = \Lambda_{1m}\bar{\epsilon}'_m, \nonumber \\
    M'_y = - M_y &=&
    -\Lambda_{21} \bar{\epsilon}_1 -\Lambda_{22} \bar{\epsilon}_2 - \Lambda_{23} \bar{\epsilon}_3 - \Lambda_{24} \bar{\epsilon}_4 - \Lambda_{25} \bar{\epsilon}_5 -\Lambda_{26} \bar{\epsilon}_6
    \nonumber \\
    &\stackrel{!}{=}& \Lambda_{21} \bar{\epsilon}_1 + \Lambda_{22} \bar{\epsilon}_2 + \Lambda_{23} \bar{\epsilon}_3 + \Lambda_{24} \bar{\epsilon}_4 - \Lambda_{25} \bar{\epsilon}_5 -\Lambda_{26} \bar{\epsilon}_6 = \Lambda_{2m}\bar{\epsilon}'_m. \nonumber
\end{eqnarray}
From these expressions, we obtain: $\Lambda_{15} = \Lambda_{16} =0$ and $\Lambda_{21} = \Lambda_{22} = \Lambda_{23}=\Lambda_{24}=0$. Consequently, the CIM tensor is now reduced to:
\begin{eqnarray}
        \Lambda = 
    \begin{pmatrix}
        \Lambda_{11} & \Lambda_{12} & \Lambda_{13} & \Lambda_{14} & 0 & 0 \\
        0 & 0 & 0 & 0 & \Lambda_{25} & \Lambda_{26} \\
        0 & 0 & 0 & 0 & 0 & 0 
    \end{pmatrix} .
\end{eqnarray}

We then apply another symmetry operation $\{ \mathcal{T}6_{001}||3_{001} \}$ in $G_{NS}$ to further constrain the CIM tensor. Under this operation:
\begin{eqnarray}
    \{ E||3_{001} \}{\tilde{\bm \epsilon}} &=& \left(-\frac{1}{2}\tilde\epsilon_x+\frac{\sqrt{3}}{2}\tilde\epsilon_y, -\frac{\sqrt{3}}{2}\tilde\epsilon_x-\frac{1}{2} \tilde\epsilon_y, \tilde\epsilon_z\right) = (\tilde{\epsilon}'_x, \tilde{\epsilon}'_y, \tilde{\epsilon}'_z), \nonumber \\
    \{ \mathcal{T}6_{001}||E\}{\bm M} &=& \left(-\frac{1}{2}M_x-\frac{\sqrt{3}}{2}M_y, \frac{\sqrt{3}}{2}M_x - \frac{1}{2}M_y, M_z\right)=(M'_x, M'_y, M'_z). \nonumber 
\end{eqnarray}
Therefore, we have:
\begin{eqnarray}
    M'_x = -\frac{1}{2}M_x-\frac{\sqrt{3}}{2}M_y &=& -\frac{1}{2}\left( \Lambda_{11}\tilde{\epsilon}^2_x + \Lambda_{12}\tilde{\epsilon}^2_y + \Lambda_{13}\tilde{\epsilon}^2_z + 2\Lambda_{14}\tilde{\epsilon}_y \tilde{\epsilon}_z\right) - \frac{\sqrt{3}}{2}\left( 2\Lambda_{25}\tilde{\epsilon}_x \tilde{\epsilon}_z + 2\Lambda_{26}\tilde{\epsilon}_x \tilde{\epsilon}_y\right) \nonumber \\
    &\stackrel{!}{=}& \Lambda_{11} \left(-\frac{1}{2}\tilde\epsilon_x+\frac{\sqrt{3}}{2}\tilde\epsilon_y\right)^2 + \Lambda_{12} \left(-\frac{\sqrt{3}}{2}\tilde\epsilon_x-\frac{1}{2} \tilde\epsilon_y\right)^2 + \Lambda_{13} \tilde\epsilon^2_{z} + 2\Lambda_{14}\left(-\frac{\sqrt{3}}{2}\tilde\epsilon_x-\frac{1}{2} \tilde\epsilon_y\right)\tilde\epsilon_z \nonumber \\
    &=& \Lambda_{1m}\bar{\epsilon}'_{m},
    \nonumber 
\end{eqnarray}
and
\begin{eqnarray}
        M'_y =\frac{\sqrt{3}}{2}M_x - \frac{1}{2}M_y &=&
        \frac{\sqrt{3}}{2}
        \left( \Lambda_{11}\tilde{\epsilon}^2_x + \Lambda_{12}\tilde{\epsilon}^2_y + \Lambda_{13}\tilde{\epsilon}^2_z + 2\Lambda_{14}\tilde{\epsilon}_y \tilde{\epsilon}_z\right) - \frac{1}{2} \left( 2\Lambda_{25}\tilde{\epsilon}_x \tilde{\epsilon}_z + 2\Lambda_{26}\tilde{\epsilon}_x \tilde{\epsilon}_y\right)
        \nonumber \\
        &\stackrel{!}{=}&
        2\Lambda_{25}\left(-\frac{1}{2}\tilde\epsilon_x+\frac{\sqrt{3}}{2}\tilde\epsilon_y\right)\tilde\epsilon_z + 2\Lambda_{26}\left(-\frac{1}{2}\tilde\epsilon_x+\frac{\sqrt{3}}{2}\tilde\epsilon_y\right)\left(-\frac{\sqrt{3}}{2}\tilde\epsilon_x-\frac{1}{2} \tilde\epsilon_y\right) \nonumber \\ 
        &=& \Lambda_{2m}\bar{\epsilon}'_{m}.
\end{eqnarray}
% Comparing these expressions with those given by Eq.~(\ref{eq: CIM general}), 
From these expressions, we obtain: $\Lambda_{11}=\Lambda_{26}=-\Lambda_{12}$, $\Lambda_{14} = \Lambda_{25}=0$ and $\Lambda_{13} =0$.
Applying other operations do not further reduce the elements.
Finally, the CIM tensor for coplanar spin group $G_{NS} = \ ^{3_z}6/^1m^{2_x}m^{2_{xy}}m$ takes the form:
\begin{eqnarray}
            \Lambda = 
    \begin{pmatrix}
        \Lambda_{11} & -\Lambda_{11} & 0 & 0 & 0 & 0 \\
        0 & 0 & 0 & 0 & 0 & \Lambda_{11} \\
        0 & 0 & 0 & 0 & 0 & 0 
    \end{pmatrix}.
\end{eqnarray}

\begin{center}
    \begin{figure}[htbp]
        \centering
        \includegraphics[width=0.6\linewidth]{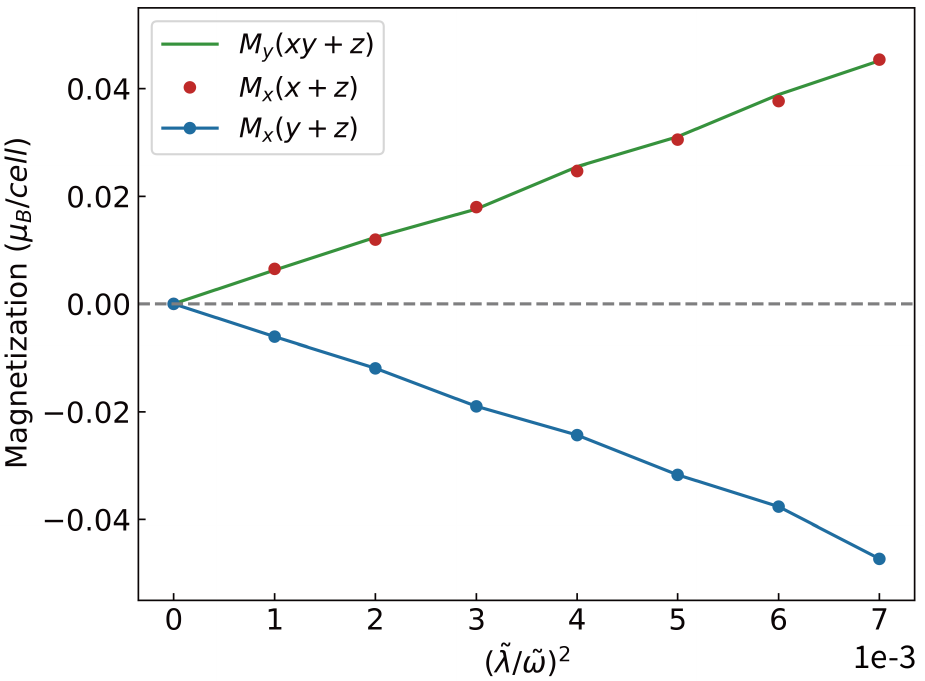}
        \caption{ Cavity-induced magnetization in AFM-1 Mn$_3$Sn without SOC.
        }
        \label{fig:Mn3Sn noSOC}
    \end{figure}
\end{center}

The induced magnetization of AFM-1 Mn$_3$Sn calculated by QEDFT without SOC is shown in Fig.~\ref{fig:Mn3Sn noSOC}. The results are fully consistent with the CIM tensor of $G_{\rm SP}$ we derived above.

% \newpage
\bibliography{main}% Produces the bibliography via BibTeX.